\documentclass[journal]{IEEEtran}
\usepackage{graphicx, mathptmx, times, epsfig, subfigure,verbatim,cite}
\usepackage{amsmath,amssymb,graphicx,latexsym,url,setspace}
\usepackage{epsfig,subfigure,mathptmx,times,latexsym,verbatim,url}
\usepackage{times,threeparttable,amsfonts,amssymb}
\usepackage{eurosym}
\usepackage{lgreek}
\usepackage{color}
\usepackage{subfigure}
\usepackage{algorithmic}
\usepackage{algorithm}
\usepackage{hhline}

\IEEEoverridecommandlockouts
\begin{document}

\title{Design of a Near-Ideal Fault-Tolerant Routing Algorithm for Network-on-Chip-Based Multicores}
\author{Costas Iordanou$^{\ast}$, Vassos Soteriou$^{\dagger}$, Konstantinos Aisopos$^{\ddagger}$\\
\begin{tabular}[t]{c@{\extracolsep{2em}}c c}
 $^{\ast}$Telematics Engineering &	$^{\dagger}$Department of EECEI & $^{\ddagger}$Google\\
Universidad Carlos III de Madrid & Cyprus University of Technology & Mountain View, CA \\
 and Telefonica I+D, Barcelona& &\\
kostas.iordanou@telefonica.com & vassos.soteriou@cut.ac.cy & kaisopos@gmail.com 
\end{tabular}}




\markboth{}
{Shell \MakeLowercase{\textit{et al.}}: Bare Demo of IEEEtran.cls for Computer Society Journals}


\maketitle
\begin{abstract}
With relentless CMOS technology downsizing Networks-on-Chips (NoCs)
are inescapably experiencing escalating susceptibility to wearout
and reduced reliability. While faults in processors and memories may
be masked via redundancy, or mitigated via techniques such as task
migration, NoCs are especially vulnerable to hardware faults as a
single link breakdown may cause inter-tile communication to halt
indefinitely, rendering the whole multicore chip inoperable. As
such, NoCs impose the risk of becoming the pivotal point of failure
in chip multicores that utilize them. Aiming towards seamless NoC
operation in the presence of faulty links we propose Hermes, a
near-ideal fault-tolerant routing algorithm that meets the
objectives of exhibiting high levels of robustness, operating in a
distributed mode, guaranteeing freedom from deadlocks, and
evening-out traffic, among many. Hermes is a limited-overhead
deadlock-free hybrid routing algorithm, utilizing load-balancing
routing on fault-free paths to sustain high-throughput, while
providing pre-reconfigured escape path selection in the vicinity of
faults. Under such online mechanisms, Hermes's performance degrades
gracefully with increasing faulty link counts, a crucially desirable
response lacking in prior-art. Additionally, Hermes identifies
non-communicating network partitions in scenarios where faulty links
are topologically densely distributed such that packets being routed
to physically isolated regions cause no network stagnation due to
indefinite chained blockages starting at sub-network boundaries. An
extensive experimental evaluation, including utilizing traffic
workloads gathered from full-system chip multi-processor
simulations, shows that Hermes improves network throughput by up to
$3\times$ when compared against the state-of-the-art. Further,
hardware synthesis results prove Hermes's efficacy.

\end{abstract}


\IEEEpeerreviewmaketitle
\section{Introduction}\label{section:introSection}

Multicore chips such as Chip Multi-Processors
(CMPs)~\cite{Intel80CoreJournal} employ Networks-on-Chips (NoCs) as
their interconnect architecture of choice to provide efficient
on-chip communication. Unfortunately, at deep sub-micron scales
on-chip components become increasingly unreliable and susceptible to
permanent faults, with the International Technology Roadmap for
Semiconductors (ITRS)~\cite{ITRSInfo} projecting a 10-fold increase
in CMOS wear rate in a 10-year span, while other studies
pessimistically forecast that in future multi-billion transistor
chips $20\%$ of the transistors will be mal-produced with a further
$10\%$ failing during their lifetime~\cite{Furber06}. While device
failure rates will keep increasing at future CMOS
technologies~\cite{Constantinescu03}, multi-billion transistor chips
containing faulty components will be expected to operate
transparently as being reliable and fault-free~\cite{DACSegmentOS}.

In CMPs, transistors found in processing cores, cache memory, and
on-chip network routers may equally fail permanently due to
time-dependent physical wearout effects such as hot-carrier
degradation and oxide breakdown~\cite{BulletProof}, that may in turn
quickly manifest to architectural-level failures. Individual core
wear-out or failures in on-chip memory modules, however, may not be
unavoidably disastrous to the CMP's full-system functionality as
cores and memory cells are inherently redundant to a particular
extent~\cite{Powell09}. Under such failure scenarios, the multicore
system may preserve its functionality, albeit at a degraded
performance mode, given that error detection and recovery
techniques, such as fault-tolerant task migration, core- and
cache-level error isolation and masking, and relevant operating
system (OS) support, are available~\cite{GizoCMPrecovery}.

NoC fabrics, however, enjoy crucially less redundancy compared to
other multicore chip components. Even an isolated intra-router
defect or a sole link failure can morph a regular topology into an
arbitrary one with an unanticipated geometry. Hence, either physical
connectivity among network nodes may not exist at all due to the
presence of faulty links and/or routers, that may even cause entire
sub-network regions to detach from each other forming distinct
partitions, and/or the associated routing protocol may not be able
to advance packets to their destinations due to protocol-level
violations. Traffic-induced back-pressure, then, causes accumulated
congestion, possible traffic stalls, and even the entire multicore
system to halt indefinitely rendering it inoperable. Hence an NoC
failure, can become the entire multicore system's \textit{single}
fatal failure.

\vspace{-3mm}
\subsection{Hermes: Synopsis and Contributions}\label{subsection:HermesIntroContributions}
\vspace{-1mm}

A key operational stress-induced wear mechanism in current and
future CMOS technologies is electromigration (EM) that causes
material deformations and consequently the loss of connections in a
circuit~\cite{Black69}. The International Technology Roadmap for
Semiconductors~\cite{ITRSInfo} assesses EM as the main cause of
on-chip metal interconnect reliability loss, an anomaly that is
amplified in high-speed/thermally-stressed NoC links.

In an effort to establish inter-router communication resilience and
hence to sustain seamless NoC operation in the presence of faulty
links, we propose Hermes\footnote{In Greek mythology Hermes was an
Olympian God, who among many roles, he was protector and patron of
travelers.}, a highly robust distributed fault-tolerant routing
algorithm that bypasses faulty network paths/regions/areas at a
gracefully performance-degrading mode with increasing faulty link
counts. Hermes employs a dual-strategy routing function that is
aware of the current topological region's health state. Upon a
message's network injection and continuing downstream in areas
encompassing healthy links only, by default Hermes utilizes either
XY Dimension-Order Routing (DOR) or load-balancing O1TURN
routing~\cite{ZeroOneTurn} to sustain high throughput. Once a faulty
link is encountered Hermes switches to using pre-calculated routing
information distributed at each router that unitedly serve in
forming an escape trail that steers that message toward its
destination. To enact the latter mode, Hermes employs a
route-discovering strategy which preempts regular operation, during
which each root-acting node broadcasts atomically route-finding
flags that spread synchronously along a spanning tree following the
Up*/Down* scheme to program said tables in all inner and leaf nodes
with how they can reach that root~\cite{Aisopos2011}. All nodes
deterministically  in turn assume this root role, so that finally
routing tables at every node contain route data on how to reach any
other NoC node.




The Up*/Down* process is tailored to provide deadlock-free routing
in a NoC fabric containing faulty links as it forms an acyclic
spanning tree that interconnects all its
nodes~\cite{DuatoDeadFreeTheory,AutoNetPaper}. Further, no switching
from Up*/Down* to either DOR or O1TURN routing is permitted in
Hermes so as to break channel dependencies and avoid inducing
deadlocks (see proof in Section~\ref{subsection:DeadlockFreeProof}).
With new faulty link(s) appearing in the topology,  current
Up*/Down*-based routes are invalidated and new ones are
reconfigured; as such the network is frozen and said
route-discovering flag-scanning process is repeated, until a new NoC
interconnectivity is calculated and all routing tables are
relevantly updated. NoC operation is then resumed. We note that
works which propose intra-router redundancy-based wear resilience,
such as in buffers and router ports~\cite{VicisGates}, equally
pivotal to an NoC's operational robustness~\cite{PehSurvey}, are
complementary and can be orthogonally applied to Hermes's scheme to
further extend an NoC's operational lifetime.

Hermes is geared towards 2D mesh NoCs, and while its Up*/Down*
faulty path discovering and routing reconfiguration process are
based on Ariadne's elegant approach~\cite{Aisopos2011}, it entails a
number of extensions and distinct contributions:

\begin{enumerate}
    \item Its \textit{two synergistic routing functions}, i.e., XY with *Up/Down*, or O1TURN with *Up/Down*,
    achieve gracefully-degrading performance with a growing faulty link count
(see Section~\ref{section:ExperimentResults}) vs.
prior-art~\cite{Aisopos2011,uDIRECpaper}.

    \item It illustrates that \textit{Virtual Channel (VC) classification} with respect to a routing path's state, i.e.,
    whether it contains healthy links only or faulty links as well, and as such usage of each VC class by a distinct pre-configured
    or pre-designed routing function, offers
    superior performance compared to the arbitrary use
of unclassified VCs, i.e., all VCs servicing a \textit{single}
routing scheme irrespective of whether in-transit messages traverse
fault-free paths or currently bypass faulty paths (see
Section~\ref{section:ExperimentResults}).

    \item It is a \textit{purely hardware-based approach} and does not employ variable execution time iterative software kernels that demand OS scheduling and
processing hardware support to discover fault-free routes,
that may interrupt multicore chip operation~\cite{uDIRECpaper}. Hermes's \textit{deterministic-time
reconfiguration} implies \textit{predictability} in network behavior
ensuring an expected level of performance attainment as
network down-time is strictly bounded~\cite{PerlmanSpanning}.

    \item Hermes \textit{identifies network segmentations} using its topology-scanning flags. The OS running on a CMP may utilize this information to mark
sub-network borders, enabling the assignment of independent
processes and threads to each network partition~\cite{DACSegmentOS}
(see Section~\ref{subsection:SubNetWalk}).
\end{enumerate}

An \textit{ideal Fault-Tolerant (FT) routing algorithm} is defined
as one that provides low-complexity all-to-all node connectivity
given a fully-connected topology, while guaranteeing deadlock- and
livelock-freedom during routing, and where its attainable throughput
degrades in direct proportion to the lost aggregated bandwidth due
to interconnect links failing. We claim that Hermes is a
\textit{near-ideal FT routing algorithm} as it meets all said goals,
however, while its sustained throughput degrades gracefully with
increasing faulty link counts (see
Section~\ref{subsection:SaturationRes}) it drops at a slightly
higher rate compared to the proportional loss in total link
bandwidth. The latter occurs as link faults may not spread evenly
across network space, or because they may lie at critical points
such as at the network bisection, and crucially, due to the fact
that in an attempt to establish freedom from deadlocks in a
decimated topology some healthy links may not be utilized by the
acyclic spanning trees that were reconfigured during Up*/Down*
marking.

\vspace{-2mm}
\subsection{Hermes's Major Features and Attributes}\label{subsection:HermesIntroCharacteristics}
\vspace{-1mm}

Being near-ideal, Hermes possesses all the following \emph{features}
and meets all design~\emph{objectives} of a well-designed FT routing
protocol~\cite{FTOddEven}, along with~\textit{extra attributes}
(pts. 5 - 6):

\begin{enumerate}
\item It establishes \emph{\textit{fault-tolerance}} bypassing a \emph{\textit{high number of faulty links}} that can form any fault region (set of
neighboring links) with no faulty link spatial placement restrictions and healthy link victimization to
strictly adhere to geometrical rules in achieving protocol-level deadlock-freedom vs. works such as~\cite{FTOddEven,DuatoBook}. Given a
\emph{realistic} minimally-connected path scenario, Hermes maintains \emph{\textit{feasibility}} in packet delivery given that \emph{any} physical
connectivity exits (see Section~\ref{section:HERMESAlgorithm}). Hence, Hermes \emph{\textit{adapts}} to the state of the topology, where any spatial
permutation of healthy and faulty links may exist.

\item It is \emph{\textit{deadlock-free}}, where no packets can be involved in a deadlocked situation which can halt the flow of packets and stall CMP
operation indefinitely (see Section~\ref{subsection:DeadlockFreeProof}), while establishing \emph{\textit{short routes}}, devoid of
\emph{\textit{livelocks}}.

\item It is \emph{\textit{distributed}}, where each node individually directs its packets towards their next-hop router, with no expensive global
information maintenance concerning the number and spatial
distribution of faulty links (see
Section~\ref{section:HERMESAlgorithm}).

\item It supports \textit{load-balancing} in fault-free regions to maintain high performance \textit{irrespective of application spatio-temporal traffic
behavior} (see Section~\ref{section:HERMESAlgorithm}); it does not require explicit offline path analysis when more links fail, or when new applications
are scheduled to run, to calculate specific topological bandwidth demands so as to tune channel selection to profiled traffic
patterns~\cite{DLeeBertaccoICCD15}.

\item It is \emph{\textit{lightweight}}, demanding \textit{reasonable area and power overheads}, while keeping the base pipelined wormhole router's
\textit{critical path unaffected} (see Section~\ref{subsection:Synthesis}).

\item It can handle \emph{\textit{dynamically-occurring (run-time) link faults}}; both transient and permanent faults can be handled, with Hermes geared
toward the latter category.
\end{enumerate}

Although in most previous works some of the above objectives can
conflict each other~\cite{FTOddEven,DallyBook,DuatoBook}, to our
best knowledge Hermes may be the \emph{first} FT routing algorithm
for NoCs to \emph{satisfy all of them}. Our experimental evaluation,
including utilizing traffic benchmarks gathered from full-system
chip multi-processor simulations, shows that Hermes reduces network
latency and improves throughput by up to $3\times$ when compared
against the state-of-the-art.

Next, Section~\ref{section:HERMESAlgorithm} details Hermes's network
reconfiguration scheme, routing algorithm and sub-network detection
mechanism while
Section~\ref{section:HERMESArch} presents its micro-architecture.
Following, Section~\ref{section:ExperimentResults} evaluates Hermes,
while Section~\ref{section:RelatedWork} discusses related work.
Finally, Section~\ref{section:conclusions} concludes this paper.

\vspace{-2mm}
\section{Hermes Routing Algorithm}\label{section:HERMESAlgorithm}
\vspace{-1mm}

With a NoC link(s) becoming faulty network operation is temporarily
halted and Hermes's reconfiguration process is initiated preemptively to discover
and mark all such faulty links so that they can be bypassed by
advancing messages once regular network operation resumes. Here, we
assume that all gate-level faults in an NoC router are mapped as
equivalent link-level faults~\cite{Aisopos2011}; hence gate-level
faults that render portions of a router as non-functional are
captured as link-level faults which ultimately inhibit inter-router
communication.

Following Ariadne's reconfiguration scheme~\cite{Aisopos2011}, route
discovery utilizes \textit{atomic} flag broadcasts emanating from
each node that in turn assume root status, spreading synchronously
from that root along an acyclic spanning tree. This connectivity
search treats a sub-connected network as a graph where the pattern
of said flag broadcasting conforms to Up*/Down* rules, a route
marking scheme originally utilized in local area
networks~\cite{AutoNetPaper} (LANs) and topology-irregular networks
of workstations~\cite{Sancho2004TPDS,SillaPaper}. We adopt this
Up*/Down* scheme to 2D meshes, the topology of choice in
silicon-planar NoCs~\cite{Intel80CoreJournal}, to program routing
tables found in all remaining nodes with how they can reach that
root. When this entire orchestrated process is done, with all nodes
assuming in turn root role in a time-slotted mode, the routing table
of each node contains information on how to reach all other
physically connected nodes. As such, full NoC routing connectivity
is established.

Up*/Down* is chosen as its topology-agnostic mode forms
\textit{directed acyclic spanning trees} that interconnect all NoC
nodes; route traversals devoid of such cyclic patterns set the
necessary and sufficient condition in guaranteeing
\textit{deadlock-freedom} in wormhole
networks~\cite{DuatoDeadFreeTheory,DuatoBook}. The entire path
discovery process converges in exactly $N^{2}$ cycles in an $N$-node
network, where, positively, this deterministic-time reconfiguration
leads to predictability in network behavior which then implies an
expected level of performance attainment as network down-time is
strictly bounded. NoC operation then resumes until a new faulty
link(s) appears; as such, the entire route-discovering process is
repeated to reproduce fresh routing paths relevant to the newly
generated irregular topology. Any faults in a router's datapath
which can block access to links even when being healthy, including
upstream and downstream buffers, control logic, and intra-router
crossbar connections, are regarded as an extension to the link(s)
connected to them, inevitably designating that link(s) as also being
unusable~\cite{uDIRECpaper}.

\vspace{-4mm}
\subsection{Routing in a Fault-Free Region}\label{subsection:RoutingNonFaultyTopo}
\vspace{-1mm}

Hermes combines two network routing strategies. At packet injection
in a completely fault-free topology or fault-free region, Hermes
uses either DOR-XY, dubbed as \textit{H-XY}, or partially adaptive
O1TURN~\cite{ZeroOneTurn} routing, named \textit{H-O1TURN}. Under
H-O1TURN, upon network ingress, a packet utilizes either XY or YX
routing with equal probability as a means of balancing network load.
Either of these routing schemes is followed until a faulty link is
encountered, where routing switches to pre-configured Up*/Down*
routes \textit{exclusively} utilized until the network egress port
is reached. H-XY uses 2 Virtual Channels (VCs), one for XY and the
second by Up*/Down*, while H-O1TURN uses 3 VCs, one for XY, the
second for YX, and the last for Up*/Down*. To preclude the formation
of cyclic dependencies  between said routing schemes and hence to
avoid the advent of protocol-induced
deadlocks~\cite{DuatoDeadFreeTheory}, no switching from Up*/Down*
routes to either DOR or O1TURN routing is permitted (see
Section~\ref{subsection:DeadlockFreeProof}).

Hermes utilizes two variants of Up*/Down* routing: 1) bidirectional
Up*/Down* as used in Ariadne~\cite{Aisopos2011}, where a fault in a
unidirectional link victimizes its paired opposite-directional link
as also being faulty, and 2), the upgraded Up*/Down* scheme of
uDIREC~\cite{uDIRECpaper}, forming two variants dubbed
\textit{H-uXY} and \textit{H-uO1TURN}, that marks unidirectional
faulty links independently, hence incurring no healthy link
victimization. However, as uDIREC requires iterative software
kernels to form unidirectional Up*/Down* paths, while Hermes is
purely hardware-based, we next focus on employing Ariadne's
Up*/Down* scheme~\cite{Aisopos2011}; H-uXY and H-uO1TURN are solely
used for performance comparisons in
Section~\ref{section:ExperimentResults}.

\vspace{-2mm}
\subsection{Routing in a Faulty Region: Reconfiguration Algorithm}\label{subsection:RouteFaultyTopo}
\vspace{-1mm}

A packet switches to Up*/Down* routing (from XY or O1TURN) and
occupies its associated VC, \textit{only when} it encounters a
faulty link in its path, and keeps following its rules until its
ejection to ensure deadlock-freedom (see
Section~\ref{subsection:DeadlockFreeProof}). The reconfiguration
process begins upon the detection of a new faulty link(s), at which
point the router to which this link is connected becomes the root
node (initiator). As in numerous other FT approaches, we consider
the presence of an online fault diagnosis/detection
solution~\cite{NoCAlertMicro} that can promptly warn Hermes of
erroneous NoC behavior. We also note that such
fault-diagnosis/detection mechanisms are orthogonal to Hermes's
route reconfiguration process.

\begin{figure}[t!]
    \begin{center}
        \includegraphics[width=.46\textwidth]{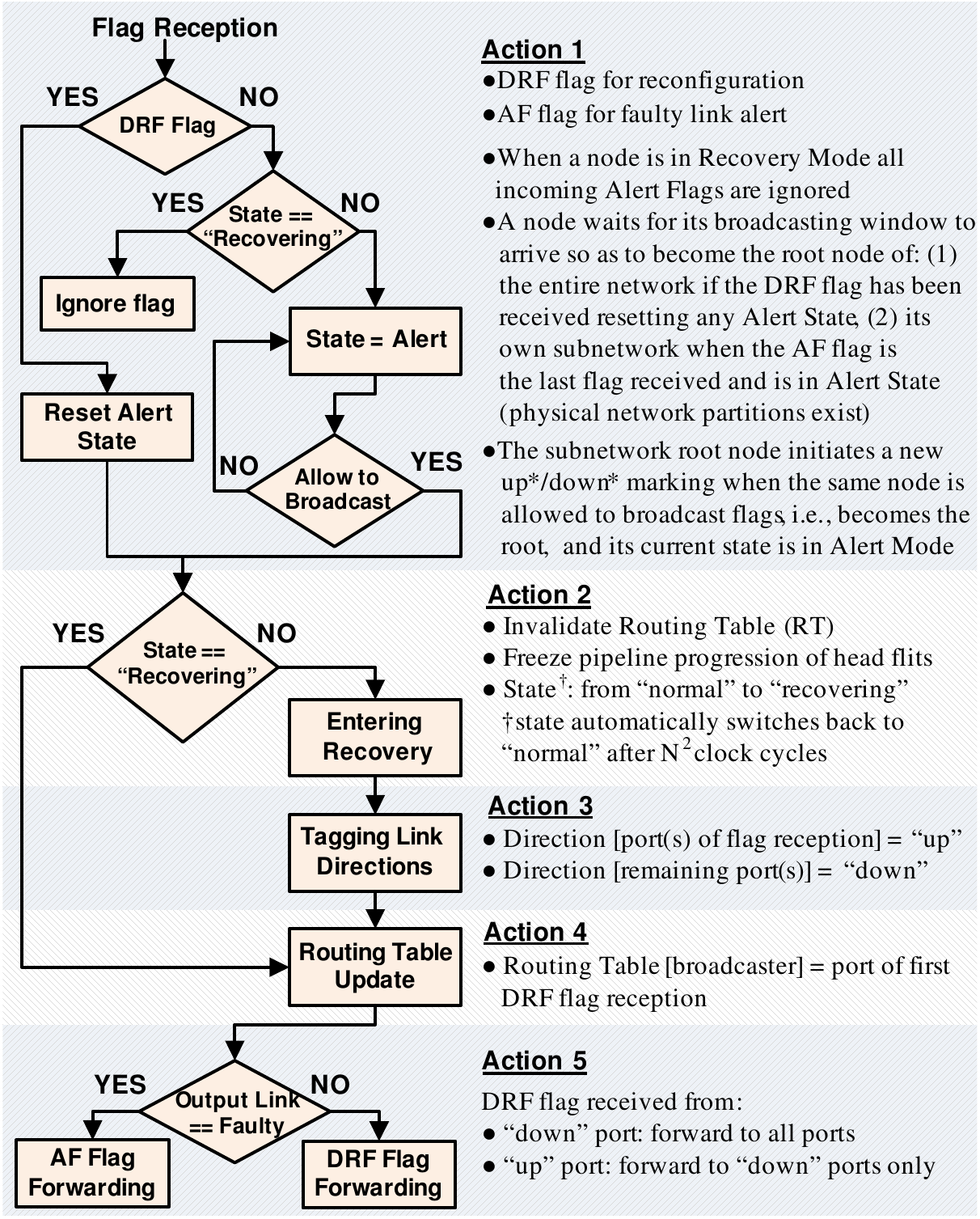}
        \vspace{-3mm}
        \caption{\small{Hermes reconfiguration algorithm.}}
        \label{fig:AlgoReconfig}
    \end{center}
    \vspace{-9mm}
\end{figure}

As such, Hermes starts broadcasting Direction-Recoding Flags
(\textit{DRF}) to all of its output ports that engage a healthy link
interconnecting to adjoining routers so as to \textit{discover the
topology's connectivity}. Flags are spread using a \textit{2-bit
overlay control network}, which sits atop the data network, to
ensure all-to-all node flag transmission/reception. The concept is
that, upon reception of these DRF flags, the next-hop neighboring
routers record the healthy input port through which the local DRF
had arrived into their routing tables so as to designate the
direction (port) that leads them back to the root (broadcasting)
node. Simultaneously (where applicable), Hermes broadcasts Alert
Flags (\textit{AF}) using the same control network to
\textit{discover network partitioning}, detailed next. Once a root
node completes reconfiguration, the remaining nodes sequentially
become root nodes, so that all nodes finally discover how to reach
each other. Each node carries the following five-stage process (see
Fig.~\ref{fig:AlgoReconfig}), requiring control information
bookkeeping and handling of reconfiguration orchestration among
routers (stage actions 1 and 5, and AF flags are unique to Hermes,
while stage actions 2-4 are inherited from
Ariadne~\cite{Aisopos2011}):

\begin{itemize}
    \item \textbf{Action 1. Flag/sub-network detection:} Identifies whether a node receives a DRF flag or an AF flag.
    \item \textbf{Action 2. Entering Recovery:} A network node enters recovery mode, invalidates its routing tables, freezes its pipeline
and stops injecting traffic into the network.
    \item \textbf{Action 3. Tagging Link Directions:} Up*/Down* marks all adjacent links of a node as either ``up'' or ``down.''
    \item \textbf{Action 4. Routing Table Update:} The routing table of the  DRF-receiving node records the cardinal direction through which the current broadcasting node is reached.
    \item \textbf{Action 5. Flag Forwarding:} Forwards the reconfiguration flag according to the link status between each node-pair (healthy=DRF flag,
faulty=AF flag). When a node is in recovering mode the AF flag is always ignored since the reconfiguration process has already begun.
\end{itemize}

When the root node completes broadcasting, the remaining $N-1$ nodes
singly broadcast their flags in a serial time-slotted mode. All five
actions are carried out during the first broadcast, while only
actions 4 and 5 are carried out during the remaining $N-1$
broadcasts. Details of each action follow next.

\textbf{Flag/Sub-Network Detection - Action 1:} The 1-bit DRF and
1-bit AF flags are broadcasted from the root; DRF flags are
propagated \emph{atop healthy links only}, while AF flags are
transmitted \emph{atop faulty links only} via the 2-bit overlay
control network (assumed to be always functional with triple modular
redundancy) interconnecting routers. The use of DRF flags is to
discover connected paths among all source-destination router pairs,
while the AF flags mark possible boundaries of network
partitions/segmentations (i.e., disjoint node clusters). As AF flags
always follow \textit{minimal} paths, while DRF flags may follow
\textit{non-minimal} paths due to the presence of faulty links,  AF
flags may arrive earlier at a router to \textit{alert} the same
router that it may possibly belong to a physically disconnected
network segment. In case a router eventually receives a DRF flag
from the root, while it had received an AF flag earlier, this
ensures that a physical path emanating from the root does exist, and
the AF flag is ignored canceling the previously set network
segmentation alert. Otherwise, the node remains in its network
segmentation alert mode (see
Section~\ref{subsection:SubNetWalk}).\vspace{-1mm}

\textbf{Entering Recovery - Action 2:} Upon DRF flag reception,
propagated from the current root node, the receiving node
invalidates its routing table, freezes flit injection, stops its
pipeline, and enters recovery. The state is set back to ``normal''
after $N^{2}$ cycles when the reconfiguration process is completed.
Each subsequent DRF flag reception will only invoke Actions 4 and 5.
However, in case the router had received an AF flag and is in its
alert state and not in recovery mode, by the end of the current $N$
cycles it will be designated that this node belongs to a network
partition. Since all nodes in all possible partitions will
eventually broadcast, all boundaries among all partitions will
eventually be discovered (see
Section~\ref{subsection:SubNetWalk}).\vspace{-1mm}

\textbf{Tagging Link Directions - Action 3:} During this step, with
the use of topology-propagating DRF flags, ports at a router are
marked as either ``up'' or ``down.'' This ensures that all
source-destination router pairs will eventually mark their routing
tables with next-hop routes to be used to reach each other. The
walk-through demo in Section~\ref{subsection:SubNetWalk} shows
how DRF flags are propagated according to pre-specified rules which
in addition maintain deadlock-free routing (see
Section~\ref{subsection:DeadlockFreeProof}).\vspace{-1mm}

\textbf{Routing Table Update - Action 4:} During each broadcast, a
root node basically informs how it can be reached from all other
nodes via the spreading of DRF flags that follows legal
Up*/Down*-based turns (see Action 5); these DRF flag broadcasting
patterns trail relevant paths starting from this root that are
recorded, in a mirror-reflected mode, into each nodes's routing
table. Any node other than the root then uses these pre-recorded
table paths to reach that root node.\vspace{-1mm}

\textbf{Flag Forwarding - Action 5:} Once the routing table of an
intermediate node is updated, the same node broadcasts its DRF flag
through those ports (via the 2-bit overlay control network) where
the associated link is healthy, that it had not received a DRF flag
from during the previous cycle, and it broadcasts an AF flag through
the port(s) where the associated link is faulty. In addition, the
Up*/Down* turn restrictions are also taken into consideration, that
is, a DRF flag received from an ``up'' link is never sent to an
``up'' link; only ``up'' to ``down'' and ``down'' to ``down''
broadcasts are legal. In case where the link is faulty, no turn
restrictions are applied, and the AF flag is sent atop the faulty
link using the 2-bit overlay  network to the neighboring
node. If the AF flag-receiving node is already in its
recovering state then the AF flag is ignored (see
Fig.~\ref{fig:AlgoReconfig}).

\vspace{-4mm}
\subsection{Timing and Synchronization}\label{subsection:TimingSync}
\vspace{-1mm}

Single-cycle flag forwarding among node pairs ensures that the
broadcasting window of each node \textit{deterministically} takes
$N$ clock cycles to complete, in an $N$-node topology. This
considers the worst-case, but unrealistic, scenario of having a
faulty topology that forms an exact \textit{minimum spanning tree},
where $N$ cycles are needed for a node's complete broadcast. Since
each node broadcasts in a non time-overlapping~\textit{atomic} mode,
reconfiguration takes $N^{2}$ cycles to complete; this scheme
establishes determinism in network behavior, implying an expected
level of performance attainment as network down-time is strictly
bounded, while the entire reconfiguration process is exactly
reproducible given a new faulty link occurrence.

\vspace{-3mm}
\subsection{Sub-Network Discovery Using AF Flags Walk-Through}\label{subsection:SubNetWalk}
\vspace{-1mm}

During network reconfiguration, a root node is identified by all
non-root nodes without explicitly forwarding its ID to them, as it
is the sole flag originator during its broadcasting window. As such,
\textit{atomic} flag broadcasts~\cite{Aisopos2011} decoded in terms of
correlating the system's global clock reference to a node's unique
ID are employed in Hermes. This math formula states that the first
$log_{2}(N)$ LSBs of the global clock designate the broadcasting
cycle of each root, while the next $log_{2}(N)$ higher bits are used
to identify the root node and hence its broadcasting slot; the
latter commands exactly $N$ repetitions (i.e., one for each
 node) using modulo arithmetic as $node((ID)mod(N))$ for node
with identity $ID$, then $node((ID+1)mod(N))$ for node with identity
$(ID+1)$ and so on until all $N$ nodes become roots independently.
Assuming an $8 \times 8$ NoC with 64 nodes, as in our experimental
evaluation of Section~\ref{section:ExperimentResults}, the first
$log_{2}(64) = 6$ LSBs and the next higher $log_{2}(64) = 6$ bits of
the global clock are used to designate the broadcasting cycle of
each node, and the root node's ID, respectively. The Node ID
Extractor (see Section~\ref{subsection:NodeIDDetector}), a logic
entity residing at every node, identifies the current root node and
uses its ID to index the Fill Routing Table Logic (see
Section~\ref{subsection:HERMESArchDetails} and
Fig.~\ref{fig:HermesLogicBlocks}-(b)) found in the same router so as
to activate the one-hot 4-bit-long routing record associated with
the current root. As such, the cardinal direction being serviced by
each of the four input ports at a router dictate the relevant ``up''
or ``down'' bit setting of this specific record dictated by
through-port-arriving DRF flags (see
Section~\ref{section:HERMESArch}). As such, every router's entire
Up*/Down*-based routing table is eventually compiled. In case
multiple concurrent faults occur, just one of the nodes becomes root
according to the arbitrary current global clock value; all other
nodes will eventually assume root status.

\begin{figure}[t!]
    \begin{center}
        \includegraphics[width=.48\textwidth]{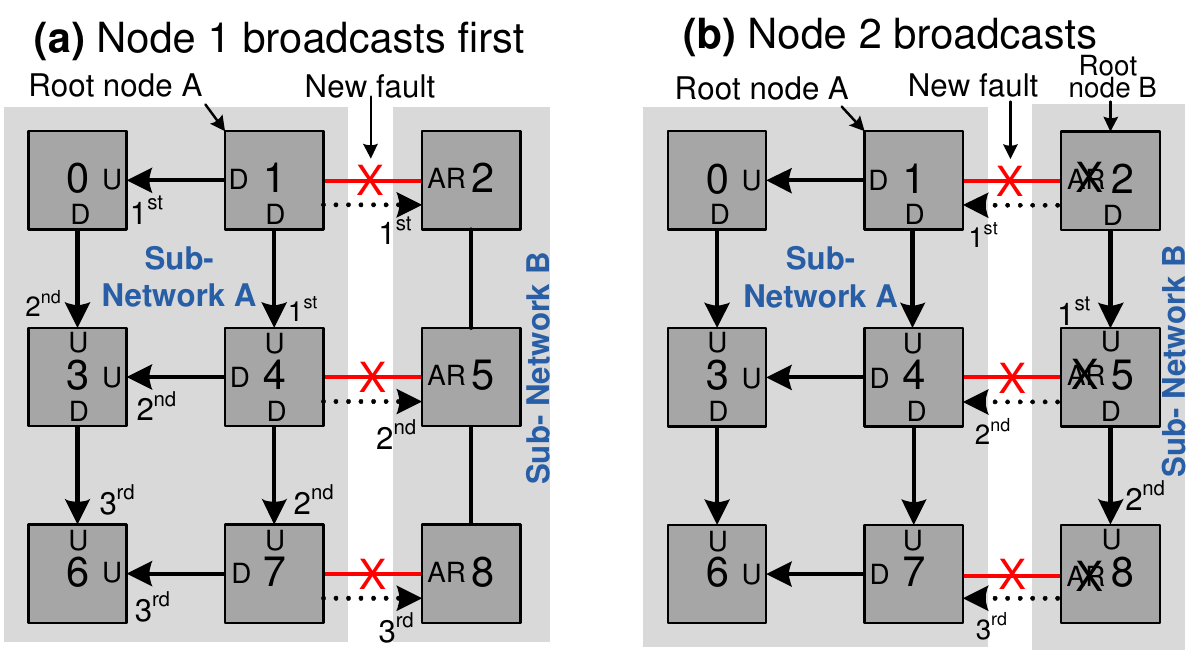}
        \vspace{-3mm}
        \caption{\small{A walk-through of Up*/Down* marking and the sub-network detection mechanism in a $3 \times 3$ mesh consisting of subnetworks A and
B. AR denotes ``Alert Register'' is on, red lines with an ``X'' are broken links, solid and dotted lines show the broadcasting of the DRF and AF flags
respectively at specified broadcast cycle times.}}
        \label{fig:SegementWalk}
    \end{center}
    \vspace{-8mm}
\end{figure}

Fig.~\ref{fig:SegementWalk} walks through Up*/Down* marking and our
Sub-Network Detection Mechanism (SNDM) that can detect individual
network-dividing non-communicating partitions, using a $3 \times 3$
mesh. The former is served by the network-spreading single-bit DRF
flags, and the latter by the single-bit AF flags. Both such flags
span the network on a cycle-by-cycle basis using the 2-bit overlay
control network (see Section~\ref{subsection:RouteFaultyTopo}).

Fig.~\ref{fig:SegementWalk} emphasizes the usage of AF flags used by
SNDM as they are a unique feature of Hermes, that traverse the
topology along with DRF flags as in Ariadne's
scheme~\cite{Aisopos2011}. In our example, links $4 \leftrightarrow
5$ and $7 \leftrightarrow 8$ are initially faulty, while link $1
\leftrightarrow  2$ presently breaks down, partitioning the topology
into sub-networks A and B. Instantly, node 1 acts as the root,
initiates the reconfiguration process, enters recovery mode, and
invalidates its current routing table entries. The broadcast cycle
times of DRF and AF flags are shown at the side of each node, which
begin spreading towards the vicinity of node 1 (see
Section~\ref{subsection:RouteFaultyTopo}). The purpose of AF flags
is to label all links residing at the physical edges of sub-networks
as border ``guards'' that mark and identify topological partitions.

As Fig.~\ref{fig:SegementWalk}-(a) shows, node 1 marks its output
ports connected to links $1 \rightarrow 0$ and $1 \rightarrow 4$ as
``down'' (D), and sends DFR flags to nodes 0 and 4. This sets their
Status Register (SR) in recovering state (see
Section~\ref{section:HERMESArch}), where their respective input
ports are set as ``up'' (U), with these port directions recorded
into their respective routing tables. Nodes 0 and 3 are aware that
node 1 is the root as the node ID is extracted from the global clock
based on modulo arithmetic (see
Section~\ref{subsection:TimingSync}), and hence all nodes are synced
(Node ID Extractor of Section~\ref{subsection:NodeIDDetector}). Node
2 only receives an AF flag, via the control network, since link $1
\leftrightarrow 2$ is now faulty, setting its Alert Register (AR) in
alert state. Following Up*/Down* rules, in cycle 2 node 0 marks link
$0 \rightarrow 3$ as D and U, and node 4 marks links $4 \rightarrow
3$ and $4 \rightarrow 7$ both as D and U, all at their two
respective ends. Nodes 3 and 7 are set in recovery state, while node
4 sends an AF flag to node 5 atop $4 \leftrightarrow 5$ faulty link,
via the 2-bit control network, setting node 5 in alert state.
Finally, in cycle 3 the north and east ports of node 6 are marked U,
setting node 6 in recovery state, while node 8 receives an AF flag
from node 7 atop $7 \leftrightarrow 8$ faulty link setting node 8 in
alert state. Now subnetwork A has completed both its Up*/Down*
marking for all of its member nodes and recording of all paths which
lead towards root node 1. Note that AF flags cause no Up*/Down*
marking at the receiving nodes, since they merely designate the
possibility of a network disconnection at the receiver node, and
that no U $\rightarrow$ U marking is allowed to ensure acyclic
behavior~\cite{Sancho2004TPDS,SillaPaper}. Also, no node that
receives a DRF or AF flag from one of its ports can later send any
other flag using that same port so as to avoid erroneous port
marking synchronization; flag reception time is tied to the global
clock dictated by the Node ID Extractor (see
Section~\ref{subsection:NodeIDDetector}).

With $N=9$ cycles elapsed, where $N$ is the node count, node 2
broadcasts next, designated by the global clock utilized by the Node
ID Extractor (see Section~\ref{subsection:NodeIDDetector}),
initiating a new Up*/Down* marking chain of events, shown in
Fig~\ref{fig:SegementWalk}-(b). The reception of an AF flag and no
DRF flag by node 2 from the previous cycle, points towards its
separate sub-network presence. Node 2 broadcasts its DRF and AF
flags accordingly, following the flag-based Up*/Down* marking rules.
The south port of node 2 is set as D while the north port of node 5
is set to U in cycle 1, and so on. However, when nodes 1, 4 and 7
receive an AF flag from corresponding nodes 2, 5 and 8 residing in
sub-network B, they ignore this flag since they are already in
recovering mode. Nodes 3-8 and 0 will broadcast in series as roots
when their turn arrives, dictated by the global clock, where each
informs its sub-network-residing nodes how they can be reached by
broadcasting DRF flags.

The process completes in $N^{2}$ cycles with node 0 ending its
broadcast. By then, \emph{only valid escape paths} are recorded into
the routing table of each router, reflecting upon routes that
\emph{are valid only within the same sub-network}, i.e., either in
sub-network A or in sub-network B. This information may subsequently
be provided to the operating system to mark the borders of these
sub-networks, enabling the assignment of independent processes and
threads to each such sub-network, being serviced by a sophisticated
algorithm such as~\cite{DACSegmentOS}.

\vspace{-2mm}
\subsection{Hermes Deadlock Freedom: Discussion \& Proof}\label{subsection:DeadlockFreeProof}
\vspace{-1mm}


We provide a short, yet comprehensive proof of deadlock-freedom
present in the H-XY and H-O1TURN routing algorithms. We adopt Duato's
theorems~\cite{DuatoFTTheory,DuatoDeadFreeTheory} which have been
instrumental in designing deadlock-free routing schemes for
mesh-based off- and on-chip interconnection networks. Duato defines
a deadlock-prone network as the one which contains cycles in its
\textit{extended channel dependency graph}. Hence, to prevent
deadlocks \textit{cyclic dependencies between channels have to be
prohibited} during routing protocol design. Next, in Fault-Tolerant
(FT) routing algorithms, connectivity, whose degree is determined
by the presentence of faults at links (physical) in combination with
the routing algorithm's reachability (by protocol), has to
considered in proving that a FT routing algorithm, such as Hermes,
is free of deadlocks.

To prove that Hermes causes no deadlocks, we first consider each
constituent routing scheme, i.e., DOR-XY (or, XY), O1TURN, and
Up*/Down* individually, and show that each is free of cyclic channel
dependencies in their context of traversed route health status. In a
second step, we then constructively prove that the combination of XY
and Up*/Down* in H-XY, and O1TURN with Up*/Down* in H-O1TURN, are
devoid of cyclic channel dependencies and hence deadlocks.


For first prove that the XY (YX) and O1TURN algorithms are
deadlock-free, despite being trivial in doing so. First XY makes
turns from any horizontal to any vertical cartesian direction,
totalling 4 turns. As such, forming overlapped abstract turns for
the clockwise and anticlockwise directions one sees that no full
cycles are possible, hence XY is devoid of deadlocks. Such analysis
applies to YX routing too. Hence, one VC is sufficient to achieve
deadlock-free DOR routing (XY or YX) in a mesh
network~\cite{DuatoFTTheory}. Next, O1TURN is implemented with two
(VCs) per physical channel (or port/link) where one VC (or
``layer''~\cite{ZeroOneTurn}) services XY and the other YX routing.
Upon packet injection, there is a $50\%$ chance in using either VC,
where switching between VCs is prohibited and a packet uses the same
VC until its network egress. Using an extended channel dependency
graph~\cite{DuatoDeadFreeTheory}, one sees that no cycles form
between the two VCs: O1TURN routing is deadlock-free.

As discussed in Section~\ref{subsection:RouteFaultyTopo} and shown
in Section~\ref{subsection:SubNetWalk}, a well-orchestrated route
discovery process executed in a distributed fashion by the augmented
logic (see Section~\ref{section:HERMESArch}) found in every NoC
router using topology-spreading flags to \textit{mark a loop-free
assignment of direction to the operational NoC links} is triggered
in the event of any link(s) becoming faulty, turning the regular
mesh topology into an arbitrary
one~\cite{SanchoNewMethodNOW,SillaPaper}. This preemptive process
fills every router's routing table with calculated routes to be used
by traversing packets in regions where all faulty links must be
bypassed, until their network egress; the algorithm is re-executed
when future faults occur to replace current routes with newly
calculated ones. The underlying principle of said link direction
assignment is the spanning tree~\cite{AutoNetPaper}, with the ``up''
(``down'') marking at each link designating the link end that is
closer (farther) to the root of the tree. The result of this ``up''
to ``down'' assignment is that the directed links, and consequently
\text{all the spanning trees} emanating from each root (source) node
to every possible leaf (destination) node do not form loops, hence
every spanning tree is acyclic. This Up*/Down* marking is guaranteed
to form a \textit{directed acyclic spanning tree} between any
root-leaf node pair, as long as there is physical connection between
them, or a series of directed healthy links. In Up*/Down* routing,
the unique node order assignments to each router, where all
increasing-order turns (down links) are disabled if followed by
decreasing-order turns (up links), or vice-versa, ensure a unique
visiting order where this decreasing to increasing node visiting
order ensures the absence of cycles, i.e., a turn from ``down'' to
``up'' is illegal and hence prevented (self-looping at a node is
also disallowed), thus eliminating full cyclic channel dependencies.
Hence, the Up*/Down* scheme is deadlock-free, for no
deadlock-producing loops are possible, as  proven
in~\cite{SillaPaper}, albeit for irregular  fault-free
networks\footnote{Link faults transform a regular mesh topology into
an irregular one, hence Up*/Down* in Hermes deals with irregular
topologies as done in~\cite{SillaPaper}.}.

Hermes's deadlock-freedom can be proven informally~\cite{SillaPaper}
by contradiction\footnote{Deadlock freedom in Hermes may be formally
proved using the step-by-step process dictated
in~\cite{DuatoDeadFreeTheory}; it is omitted here due to space
limitations.} as follows: consider that there is a deadlocked
configuration due to a cyclic channel dependency between either XY
or YX and Up*/Down* routing, as messages are allowed to switch from
either VC assigned for XY or YX usage to the VC belonging to
Up*/Down*, and then back to the VC belonging to either XY or YX. In
this configuration, messages cannot have switched to either XY or YX
routing and associated VCs if the same messages had occupied the VC
belonging to Up*/Down* routing as Hermes prohibits such a channel
switch. Thus, all blocked messages must currently be occupying a VC belonging to
either YX or XY. However, those protocol-deadlocked messages can
bypass a faulty link lying in their minimal path, which would
cause physical blockage, by using the VC belonging to Up*/Down*, as
Up*/Down* routing is by design deadlock-free, bypasses faulty links,
and can deliver those messages to their destinations given lack of
network partitioning. Hence, there cannot be a deadlocked
configuration, both in terms of protocol effects and physical
blockage due to the presence of a faulty link(s). As such, both H-XY
and H-O1TURN Hermes variants are deadlock-free.

\vspace{-2mm}
\subsection{Maximum Number of Faulty Links Handled by Hermes}\label{subsection:MaxFaultHandled}
\vspace{-1mm}

We model a NoC as a strongly connected directed mesh graph $G=(V,
C)$ with node (router) set $V$ and edge (link) set $C$ with a total
number of edges $|C| = \sum_{i=0}^{n-1} c_{i}$, where all such edges
possess healthy status. A \textit{minimum spanning tree} $S=(V,C')$
is an acyclic and connected sub-graph of $G$ such that it connects
\textit{all} the \textit{distinct} nodes in $G$ via a path of finite
sequence of edges of non-faulty status according to the Up*/Down*
scheme marking rules in the form $v_{0} \rightarrow v_{1}
\rightarrow \dotsc ,\rightarrow v_{m-2} \rightarrow v_{m-1}$ such
that $c_{0} \neq c_{m-1}$, i.e., $S$ contains no cycles. Also, $n >
m$ so that $C' \subset C$. Let $F$ be the set of faulty edges in $G$
so that $C' \cup F = \emptyset$ and $C = C' \cup F$. As such, the
maximum number of faults that can be handled by Hermes is $|F| = |C|
- |C'|$. For example, in an $8 \times 8$ mesh NoC, used in our
experimental evaluation in Section~\ref{section:ExperimentResults},
$|C| = 112$ and $|C'|=63$; hence, $|F|=49$ faulty bidirectional
links (or, $43.8\%$) can be handled by Hermes at maximum with
guaranteed packet delivery among all routers in a deadlock-free
mode.


\vspace{-2mm}
\subsection{Breadth-First Search (BFS) vs. Depth-First Search (DFS)}\label{subsection:BFSvsDFS}
\vspace{-1mm}

\begin{figure}[t!]
    \begin{center}
        \includegraphics[width=.48\textwidth]{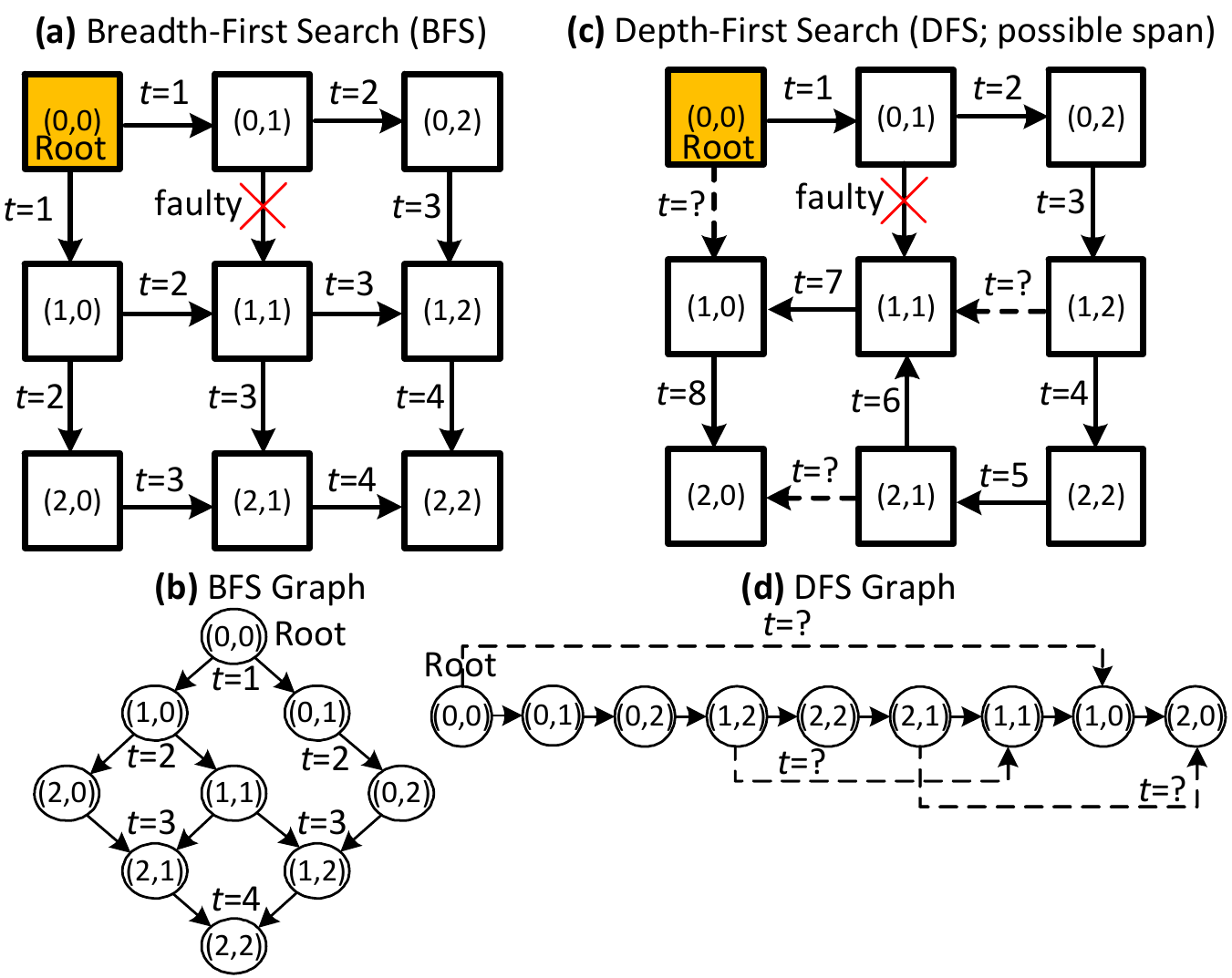}
        \vspace{-3mm}
        \caption{\small{A comparison of BFS vs DFS algorithms applied to a $3 \times 3$ mesh NoC.
        $t$ indicates the search cycle where ``?'' means unknown.}}
        \label{fig:BFSvsDFS}
    \end{center}
    \vspace{-8mm}
\end{figure}

Hermes utilizes BFS to scan the interconnect graph during Up*/Down*
marking using DRF flags to build routing tables, as
Fig.~\ref{fig:BFSvsDFS}-(a) shows, where link $(0,1) \leftrightarrow
(1,1)$ is faulty. BFS is very efficient when applied to regular
topologies, e.g., meshes, due to the algorithm's fork-like spreading
nature; as Fig.~\ref{fig:BFSvsDFS}-(b) shows, it takes 4 cycles to
search the entire graph starting from the root with cartesian
coordinates (0,0). DFS can be highly inefficient when used in regular
topologies as its search spreads linearly, requiring much more time
to build a graph; as Fig.~\ref{fig:BFSvsDFS}-(c) shows, when DFS is
applied to the same mesh topology it takes 8 cycles to build the
graph backbone (solid links in Fig.~\ref{fig:BFSvsDFS}-(d)), while
it recurses in order to build edges in the DFS graph that capture
still undiscovered links, i.e., those dotted in
Fig.~\ref{fig:BFSvsDFS}-(d). As this is highly-dependent on the
root's location~\cite{SanchoFlex}, the exact convergence time is
highly variable.

A major issue with Up*/Down*, however, is that depending on the
spatial distribution and number of faults, it creates non-minimal
routes in the quest of avoiding cyclic channel dependencies and
hence deadlocks  (see Section~\ref{subsection:DeadlockFreeProof}).
However longer paths point to an increase in the mean hop count of
packets, hence latency is increased and the network's effective
throughout is reduced. For these exact reasons, various works
applicable to the domain of off-chip Networks of Workstations (NoWs)
proposed either, (a) the relaxation of Up*/Down* in searching for
routes and the later use of heuristics as inputs to iterative
methods so as to break cycles~\cite{Sancho2004TPDS,SillaPaper}, or (b),
the use of DFS in irregular NoWs in order to enact flexibility
in constructing connectivity graphs so as to introduce previously
unconsidered legal connections among routers that improve
performance, by up to $3\times$, with shorter
routes~\cite{SillaPaper,SanchoFlex}.

Both such methods, however, exhibit performance gains that are
highly dependent on the choice of initial root node, are applicable
to non-faulty networks, and require use of heuristics in tandem with
recursive search, including trying out alternative root nodes during
tree search. The latter two are implemented using augmented hardware
that demand high overhead investment and/or using software tools and
runtime OS  kernels that are computationally expensive.

Despite the fact that Hermes deals with an irregular topology once a
first link fault occurs, we argue that such said methods should not
or cannot be applied to Hermes as they negate its very purpose:
Hermes is a  lightweight, purely hardware-based scheme suitable for
resource-constrained NoCs, it does not require recursive software
methods to discover routes and it is thus computationally oblivious,
it rotationally designates all nodes as acting roots when performing
Up*/Down* graph search, hence no optimization in root choice may be
applied, and given its near-ideal FT routing nature it dictates
deterministic-time reconfiguration  (see
Section~\ref{section:HERMESAlgorithm}). Yet, Hermes is flexible as
it can be used in software-based iterative methods, demonstrated
with our construction and testing of H-uXY and H-uO1TURN scheme
variants (see Section~\ref{section:ExperimentResults}).

\vspace{-2mm}
\section{Hermes Micro-Architecture}\label{section:HERMESArch}
\vspace{-1mm}

\begin{figure}[t!]
    \begin{center}
        \includegraphics[width=.49\textwidth]{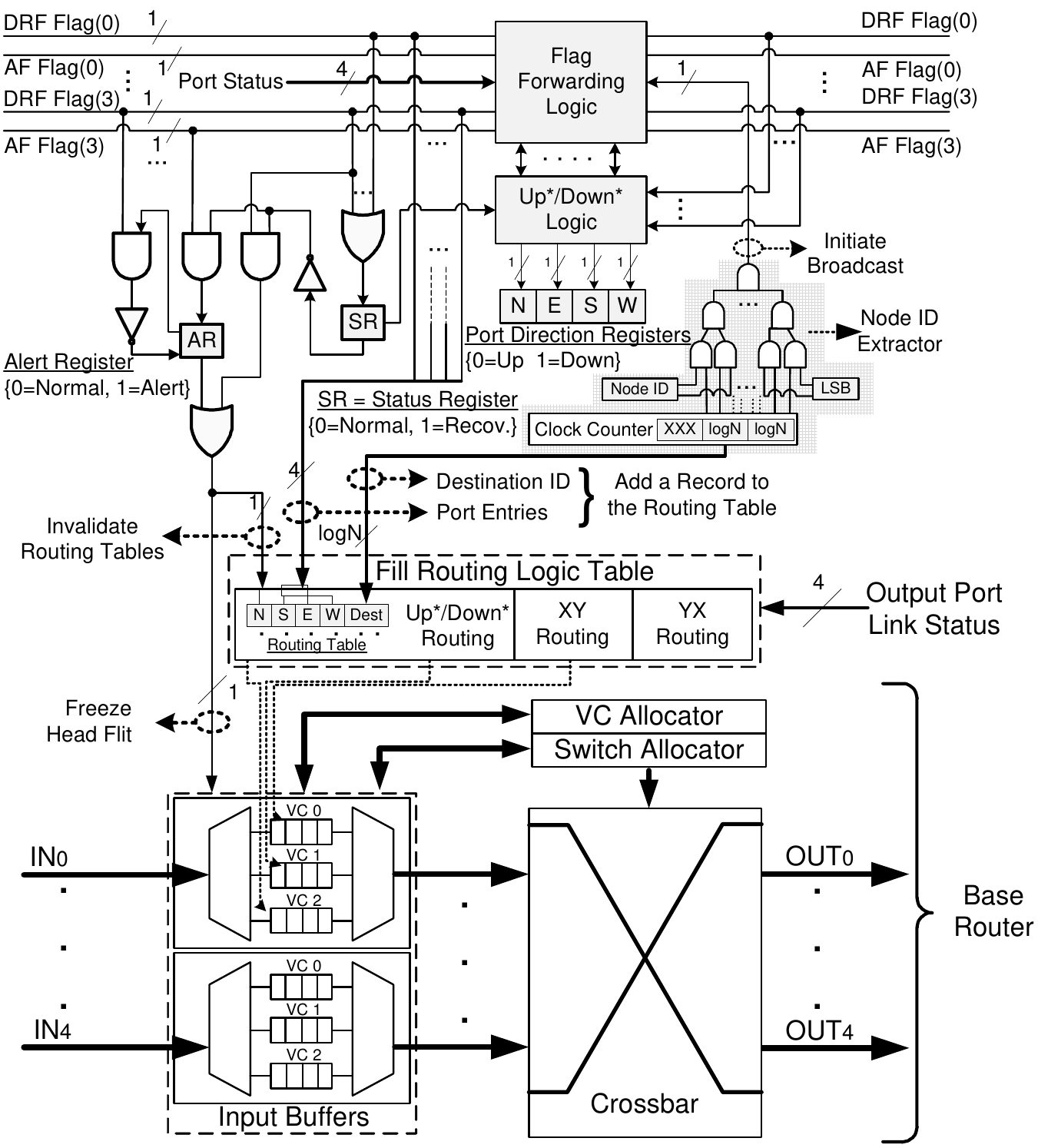}
        \vspace{-7mm}
        \caption{\small{Schematic of an input-buffered wormhole flow-control router architecture with virtual channels incorporating Hermes components.}}
        \label{fig:HermesRouterArch}
    \end{center}
    \vspace{-8mm}
\end{figure}

Hermes uses a 4-stage pipelined wormhole NoC router with 3 input
Virtual Channels (VCs), comprising (1) routing computation, (2) VC
arbitration, (3) switch allocation, and (4) crossbar
traversal\footnote{Hermes is orthogonal to other existing pipelined
NoC router architectures such as the speculative wormhole
architecture in~\cite{PehDelayModel}.}. Inter-router link traversal
consumes 1 cycle. Fig.~\ref{fig:HermesRouterArch} depicts Hermes's
router micro-architecture that incorporates added circuitry required
for Hermes's operation.

Hermes's Fill Routing Logic Tables unit, shown in
Fig.~\ref{fig:HermesRouterArch}, is divided into three sub-blocks,
comprising Up*/Down*, DOR-XY and DOR-YX routing, each utilizing a
separate VC. H-XY routing utilizes the first two blocks, while
H-O1TURN utilizes all three blocks where XY or YX routing are used
with equal probability when routing in a fault-free region. The
Up*/Down* block is directly connected to Hermes's logic unit which
provides next-hop routing information, as
Fig.~\ref{fig:HermesLogicBlocks}-(b) shows, stored in routing
tables, and is employed when the current packet being routed has
encountered a faulty link in its progressive path. These routing
tables are updated every time a new faulty link(s) appears in the
network, dictated by the route reconfiguration algorithmic process
described in Section~\ref{subsection:RouteFaultyTopo}. The VC
Allocator assigns the sole VC governed by Up*/Down* routing to a
packet being routed in a faulty region, up until that packet reaches
its destination to maintain deadlock-freedom (see
Section~\ref{subsection:DeadlockFreeProof}); otherwise XY's VC is
assigned under H-XY, or either of the XY or YX VCs are utilized in
H-O1TURN when traversing a fault-free region.

Fig.~\ref{fig:HermesRouterArch} shows Hermes's various logic
components comprising the 1-bit Status Register (SR) with ``normal''
and ``recovering'' states, the 1-bit Alert Register (AR) with
`normal'' and ``alert'' states, four 1-bit registers to designate
``up'' or ``down'' routing (one for each port) when Up*/Down* is in
use, the logic to update the SR and AR registers accordingly (Action
3; see Section~\ref{subsection:RouteFaultyTopo}), to fill the
routing tables (Action 4), and to forward the DRF or AF flags
(Action 5). Hermes's router has sixteen 1-bit width
wires for flag reception and forwarding, shown as $flag_{0-3}(in)$
and $flag_{0-3}(out)$ for flag in and out signals.
Each $flag(in)$ or $flag(out)$ pair at each port consists of two
1-bit wires, one for each DRF and AF flag forwarding.

The SR holds the current state of the network, where ``recovering''
implies that the network is currently stalled to reconfigure the
routing tables, and where ``normal'' implies that the network is
operating normally or it has resumed normal operation after
completing its routing table reconfiguration process  as a response
to fault(s) recovery. Under the normal state packets can be injected
into the network since valid routing paths exist in the routing
table of all NoC routers. The AR is updated as follows: if a new AF
flag is received and the SR is in the ``normal'' state, then the AR
register is set to the ``alert state.'' If an AF flag is received
under a current recovering state, then the AF flag is ignored. If
the AF flag had already been received and the AR register is
currently in the ``alert'' mode, then if a DRF flag is also
received, the AR is reset to the ``normal'' state (see
Section~\ref{subsection:RouteFaultyTopo} and
Section~\ref{subsection:SubNetWalk}).

\vspace{-2mm}
\subsection{Hermes's Router Logic Blocks}\label{subsection:HERMESArchDetails}
\vspace{-1mm}

\begin{figure*}[t!]
\begin{center}
\includegraphics[width=.99\textwidth]{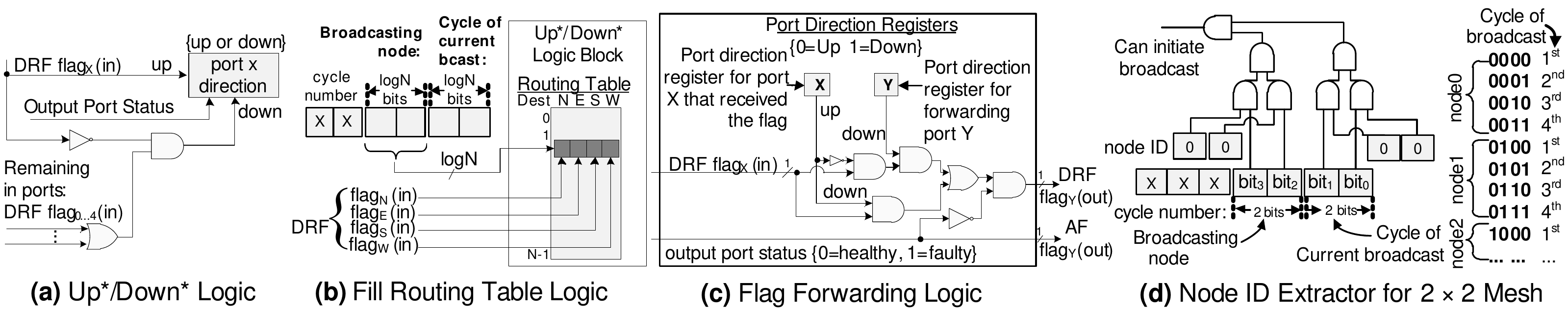}
\vspace{-3mm} \caption{\small{Schematics of all Hermes's router
logic blocks.}}
\label{fig:HermesLogicBlocks}
\end{center}
\vspace{-8mm}
\end{figure*}

When the SR is in recovering state, the signal which freezes head
flits and invalidates the routing tables is set to ``on.'' This
informs the Up*/Down* routing table block
(Fig.~\ref{fig:HermesLogicBlocks}-(a)) to discard the information
currently held in it. The input buffers and injection ports are also
stopped from forwarding/injecting packets into the network. During
this recovering state and the initial DRF flag reception at each
node in the network, the Up*/Down* logic block is utilized by the
root node and the subsequent nodes to mark their output and input
ports as either ``up'' or ``down'' according to the route
reconfiguration rules detailed in
Section~\ref{subsection:SubNetWalk}. The flag forwarding logic and
output port status signals (Fig.~\ref{fig:HermesLogicBlocks}-(c))
co-operate to determine when and what kind of flag (DRF or AF) needs
to be forwarded to each direction according to (1) the Up*/Down*
logic values and marking scheme (refer to
Section~\ref{subsection:SubNetWalk}), and (2) link statuses (refer
to Section~\ref{subsection:RouteFaultyTopo}).

The four input 1-bit flag wire signals (see
Fig.~\ref{fig:HermesLogicBlocks}-(c)) each direct the Up*/Down*
marking in their respective cardinal direction (N, S, E, or W). The
output port status identifies the health state of each output port;
in case the port is non-healthy, i.e., faulty, then the AF flag is
forwarded to the adjoining node, otherwise the DRF flag is forwarded
(see Section~\ref{subsection:SubNetWalk}). The DRF flags are
received by the routing table filling logic
(Fig.~\ref{fig:HermesLogicBlocks}-(b)), and along with the
extraction of the broadcasting node's ID and the direction of the
input port(s) from which the DRF flag(s) was received, the
corresponding entry in the routing table is updated using one-hot
encoding to designate the cardinal direction. The table consists of
$N$ 4-bit entries. An entry is recorded once per route
reconfiguration process and may only be re-recorded when a later
faulty link is detected in the NoC which initiates a fresh route
re-discovery process.


\vspace{-2mm}
\subsection{Node ID Extractor}\label{subsection:NodeIDDetector}
\vspace{-1mm}

The Node ID Extractor, depicted in
Fig.~\ref{fig:HermesRouterArch}, comprises the hardware which
identifies the currently broadcasting node during a route
reconfiguration process. It uses a set of $2log_{2}(N)$ bits, where
$N$ is the number of nodes in the NoC, with its $log_{2}(N)$ first
LSBs being used to sync to the current $log_{2}(N)$ lower bits of
the global clock so as to designate the broadcasting cycle of each
node, while the  $log_{2}(N)$ next higher bits are used to identify
the flag-broadcasting node (see
Section~\ref{subsection:SubNetWalk}). As a demonstrating example,
Fig.~\ref{fig:HermesLogicBlocks}-(d) presents such hardware (see
Section~\ref{subsection:TimingSync}), assuming a four-node $2 \times
2$ mesh network. The first two ($log_{2}(N=4)=2$) Least Significant
Bits (LSBs) indicate the broadcasting cycle number, and the next 2
bits indicate which node is broadcasting during the current
topology-scanning flag transmission window. For each individual node
the node ID is unique (Fig.~\ref{fig:HermesLogicBlocks}-(d)); hence,
when the cycle number equals to the hard-coded node ID, then that
node is allowed to broadcast, i.e., nodes 0 and 1 each broadcast
when $bit\{3,2\} = \{0,0\}$ and $bit\{3,2\}=\{0,1\}$, respectively,
and so on. This process continues until all nodes finally broadcast
their flags during a route reconfiguration process which lasts for
$N^{2}$ cycles at max (see
Section~\ref{subsection:RouteFaultyTopo}).

\vspace{-2mm}
\section{Experimental Setup and Results}\label{section:ExperimentResults}
\vspace{-1mm}

To evaluate Hermes's performance we implemented a detailed
cycle-accurate simulator that supports 2D meshes with four-stage
pipelined routers (see Section~\ref{fig:HermesRouterArch}), each
with 1 to 3 VCs per input port each consisting of 5-flit buffers.
Our framework utilizes (a) synthetic Uniform Random (UR) traffic
where all nodes have an equal probability of sending/receiving a
packet per unit time, (b) Transpose (TR) traffic where packet source
and destination coordinates matrix-alternate, an adversarial
NoC-stressing form of traffic, and (c) real application workloads.
The latter traffic traces are gathered from multi-threaded
application execution in a full-system simulation environment,
specifically from the Netrace benchmark suite~\cite{Netrace2010},
with their packets maintaining dependencies among them, tracked by
our simulator for accuracy and fidelity. Synthetic traces, with
six-flit 128-bit packets, are run for a million clock cycles, while
in Netrace applications results were gathered within a 150-million
cycle ``region of interest.'' All experiments were run using an $8 \times
8$ 2D mesh NoC topology.

Next, two distinct faulty link spatial topology-placing scenarios
were utilized in tandem with said traffic patterns: (a) fully
random, where all link faults are purely randomly distributed in the
NoC's topology, and (b) Hotspot (HS), where half of the faulty links
are randomly placed within a $4 \times 4$ 2D sub-mesh area centrally
positioned within the $8 \times 8$ 2D mesh topology, covering $25\%$
of the NoC's space; the rest of the interconnect faults are randomly
mapped onto the remaining network links outside this ``hotspot''
area. Hence, the hotspot area exhibits a 4$\times$ faulty links
density vs. the peripheral NoC area encompassing the rest of the 48
routers, purposely stressing Hermes's robustness capability in
detrimental conditions.

Under both faulty link distributions, full topology connectivity was
maintained. For fairness, 50 experiments were repeated for each
simulation point, to even-out the idiosyncrasy of each individual
spatial fault placement, with results averaged. As stated in
Section~\ref{subsection:HermesIntroContributions}, gate-level faults
that render portions of a router as non-functional excite relevant
link-level faults. A range of network faulty link counts was
considered, up to a severe $>12\%$ faulty NoC links; though such
scenario is extreme, failure probabilities have only been
\textit{predicted} for some circuit primitives in future aggressive
CMOS technology nodes~\cite{ITRSInfo}, hence the only scope here is
to stress-test Hermes. We compare Hermes against
Ariadne~\cite{Aisopos2011} and uDIREC~\cite{uDIRECpaper} which also
utilize Up*/Down* routing in NoCs, with both shown to outperform advanced FT
schemes such as Immunet~\cite{ImmunetPaper} and
Vicis~\cite{VicisGates}. We note that Ariadne's syntectic traffic
pattern results reported in the original paper~\cite{Aisopos2011}
may differ slightly from our results (see
Section~\ref{subsection:RandomPlaceFaultsRes}) due to idiosyncrasies
inherent in link fault placements and traffic patterns.

\vspace{-3mm}
\subsection{Results With Random Faulty Link Placement}\label{subsection:RandomPlaceFaultsRes}

\begin{figure}[t!]
    \begin{center}
        \includegraphics[width=.50\textwidth]{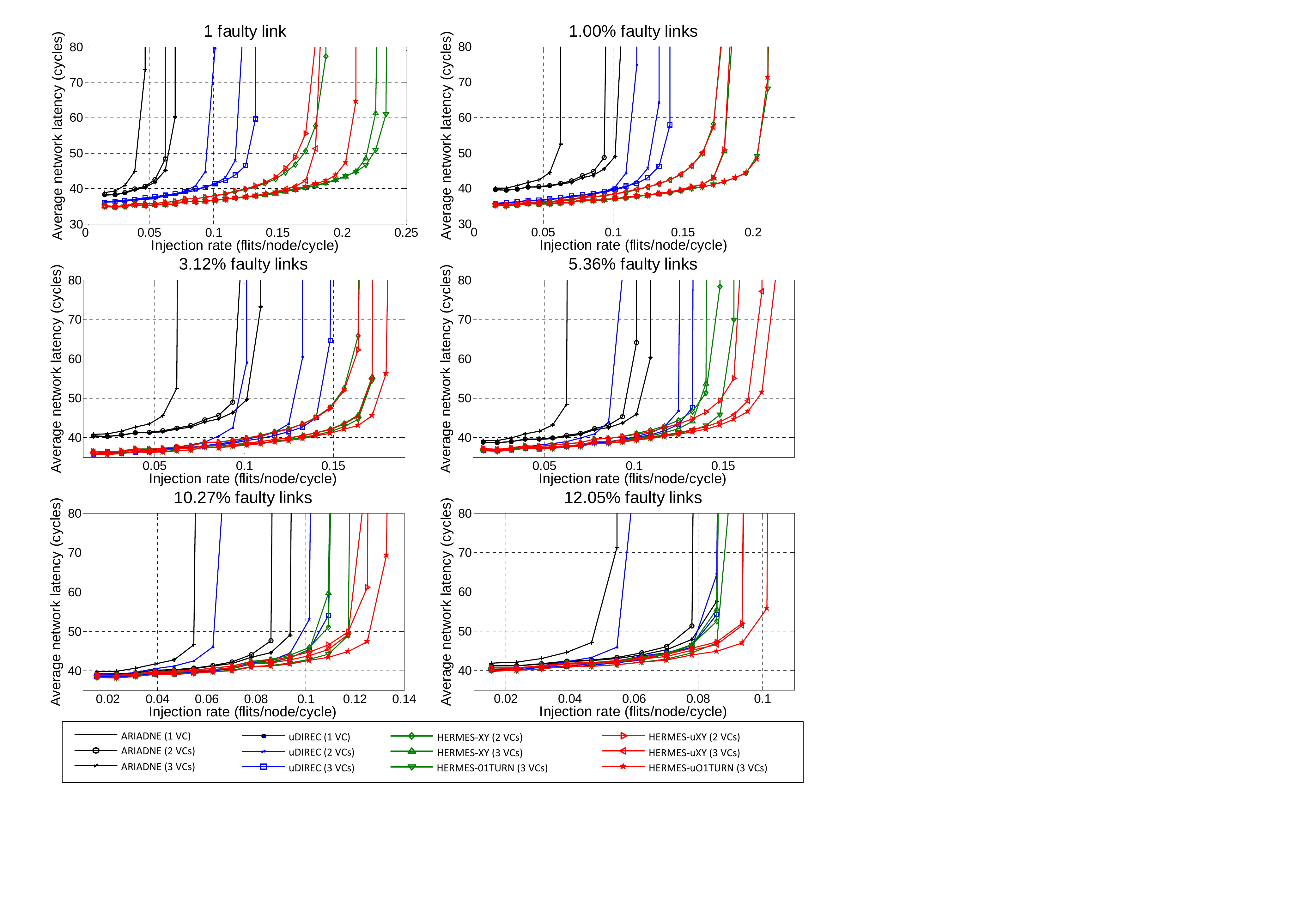}
        \vspace{-7mm}
        \caption{\small{Latency-throughput curves under a fully random faulty link placement scenario with synthetic uniform random traffic.}}
        \label{fig:UniRandPlaceUniRes}
    \end{center}
    \vspace{-8mm}
\end{figure}

Fig.~\ref{fig:UniRandPlaceUniRes} shows latency-throughput results
under UR traffic, using 1 to 3 VCs per port for all considered
routing schemes. Under all faulty link density scenarios, all Hermes
variants outperform all 3 Ariadne variants (i.e., 1-3 VCs per port)
due to the fact that Ariadne's heavy victimization of healthy
bidirectional links in enforcing Up*/Down* routing rules in pursue
of deadlock-freedom compromises NoC performance.

For up to $5.36\%$ faulty links in the topology all Hermes variants
with 2 VCs per port outperform uDIREC with 3 VCs per port. This
shows that \textit{VC classification} in terms of a routing path's
state, i.e., whether it contains healthy links only (i.e., Hermes
uses either XY or O1TURN) or faulty links as well (i.e., Hermes uses
Up*/Down* only), is \textit{superior when compared to the arbitrary
use of unclassified VCs by a single routing scheme} (i.e., the
exclusive use of Up*/Down* by either Ariadne or uDIREC) irrespective
whether in-transit messages traverse healthy paths or bypass faulty
paths. Under $10.27\%$ or $12.05\%$ faulty links, H-XY with 2 VCs
performs equally to uDIREC using 3 VCs. Evidently, under UR traffic,
the performance of all six Hermes variants degrades gracefully in
terms of the achieved saturation throughput with increasing faulty
links count. Further, with just one faulty link present in the
entire topology, its negative performance impact under Hermes is
constrained, as opposed to the cases of using Ariadne or uDIREC
where the attained saturation throughput is reduced dramatically.
Indicatively, under $5.36\%$ faulty links H-XY and H-uXY, with each
utilizing 2 VCs per port, outperform Ariadne and uDIREC, again with
each utilizing 2 VCs per port, by $39.6\%$ and $12.8\%$, and by
$56.4\%$ and $26.4\%$, respectively. Under the same faulty links
setup, H-XY and  H-uXY with each utilizing 3 VCs per port, exceed
Ariadne and uDIREC, again with each utilizing 3 VCs per port, by
$28.7\%$ and $7.2\%$, and by $49.6\%$ and $24.6\%$, respectively,
while H-O1TURN outperforms Ariadne with 3 VCs by $35.7\%$ and
H-uO1TURN tops uDIREC wth 3 VCs by $31.2\%$.

\begin{figure}[t!]
\begin{center}
\includegraphics[width=.49\textwidth]{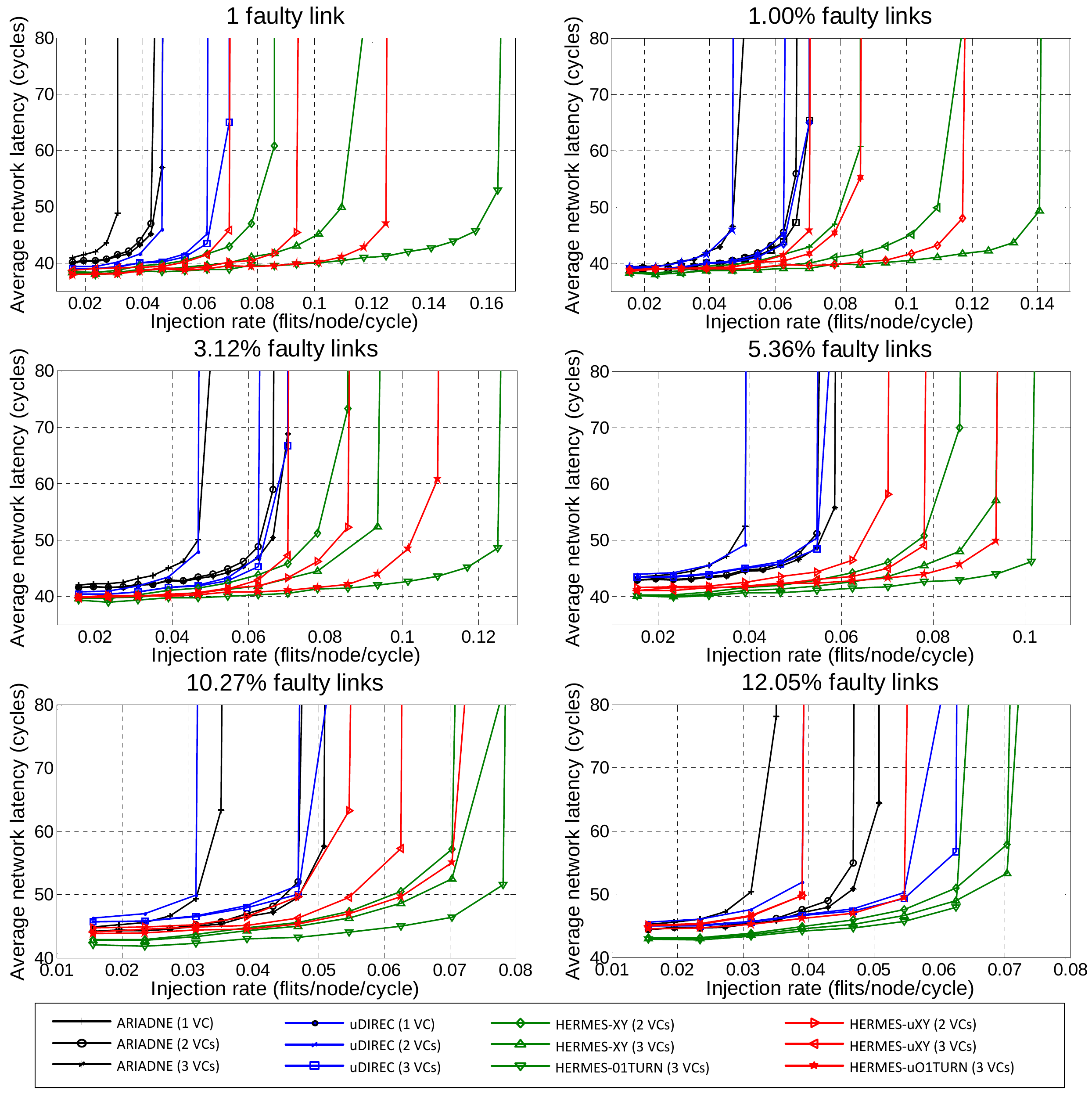}
\vspace{-7mm} \caption{\small{Latency-throughput curves under a
fully random faulty link placement scenario with synthetic transpose
traffic.}} \label{fig:UniRandPlaceTransRes}
\end{center}
\vspace{-8mm}
\end{figure}

Fig.~\ref{fig:UniRandPlaceTransRes} shows latency-throughput results
under TR traffic, using 1 to 3 VCs per port for all examined routing
schemes. Again, the performance of all six Hermes variants degrades
gracefully in terms of the achievable saturation throughput with
increasing faulty links count, and they outperform Ariadne and
uDIREC under all equivalent faulty link count scenarios and per-port
VC counts. It is also interesting to see that the non-uDIREC Hermes
variants (e.g., H-XY with 3 VCs) actually perform better than the
Hermes uDIREC variant counterparts (e.g., H-uXY with 3 VCs); this is
due to the fact that the pre-determined routes under uDIREC do not
suit non-uniform spatially unbalanced forms of traffic, which
overload the network's virtual diagonals, such as transpose traffic
which is utilized here. Indicatively, under $5.36\%$ faulty links,
H-uXY (2 VCs) and H-uO1TURN (3 VCs per port) exceed Ariadne and
uDIREC, both with 3 VCs, by $41.4\%$ and $57.7\%$, and $58.6\%$ and
$76.9\%$, respectively. Next, under $10.27\%$ faulty links, H-uXY (2
VCs per port) and H-uO1TURN (3 VCs per port) top Ariadne and uDIREC,
with each utilizing 3 VCs per port, by $5.9\%$ and $16.1\%$, and
$39.2\%$ and $52.7\%$, respectively.

\vspace{-3mm}
\subsection{Results with Hotspot Faulty Link Placement}\label{subsection:fictyfiftyHotspotRes}
\vspace{-1mm}

The hotspot spatially-distributed faulty link scenario crowds
unhealthy links in the topology center to purposely stress Hermes's
robustness capability, as the vertical and horizontal network
bisections, which convey most of the mesh's traffic, overlap
perpendicularly in the topology middle. This is even more
critical in the case of TR traffic in which all messages traverse
the two virtual network diagonals which overlap at the very center
of the topology, penetrating the hotspot area.

\begin{figure}[t!]
\begin{center}
\includegraphics[width=.51\textwidth]{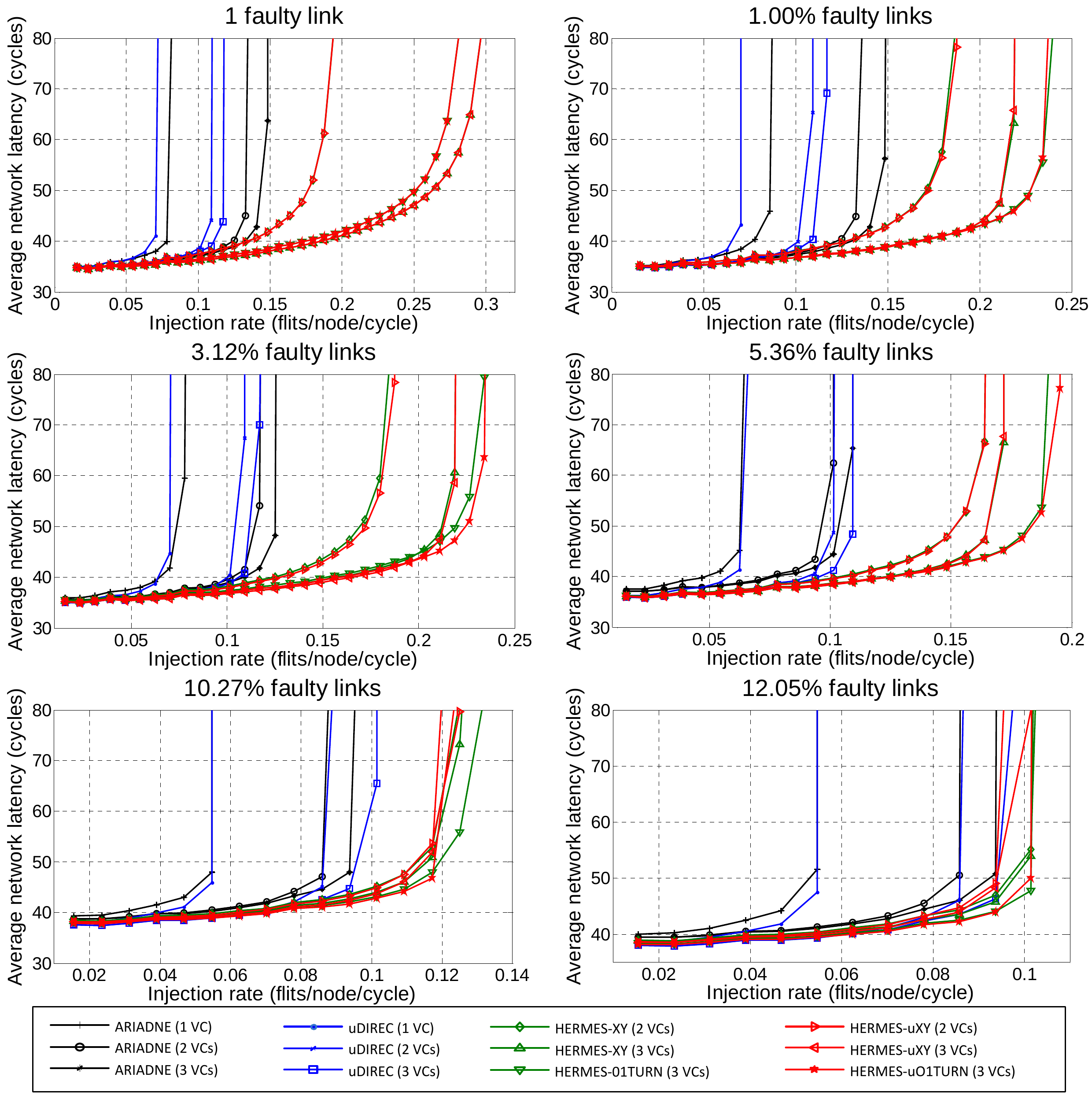}
\vspace{-6mm} \caption{\small{Latency-throughput curves under the
hotspot faulty link placement scenario with synthetic uniform random
traffic.}} \label{fig:Hotspot50UniformResults}
\end{center}
\vspace{-8mm}
\end{figure}

\begin{figure}[t!]
\begin{center}
\includegraphics[width=.49\textwidth]{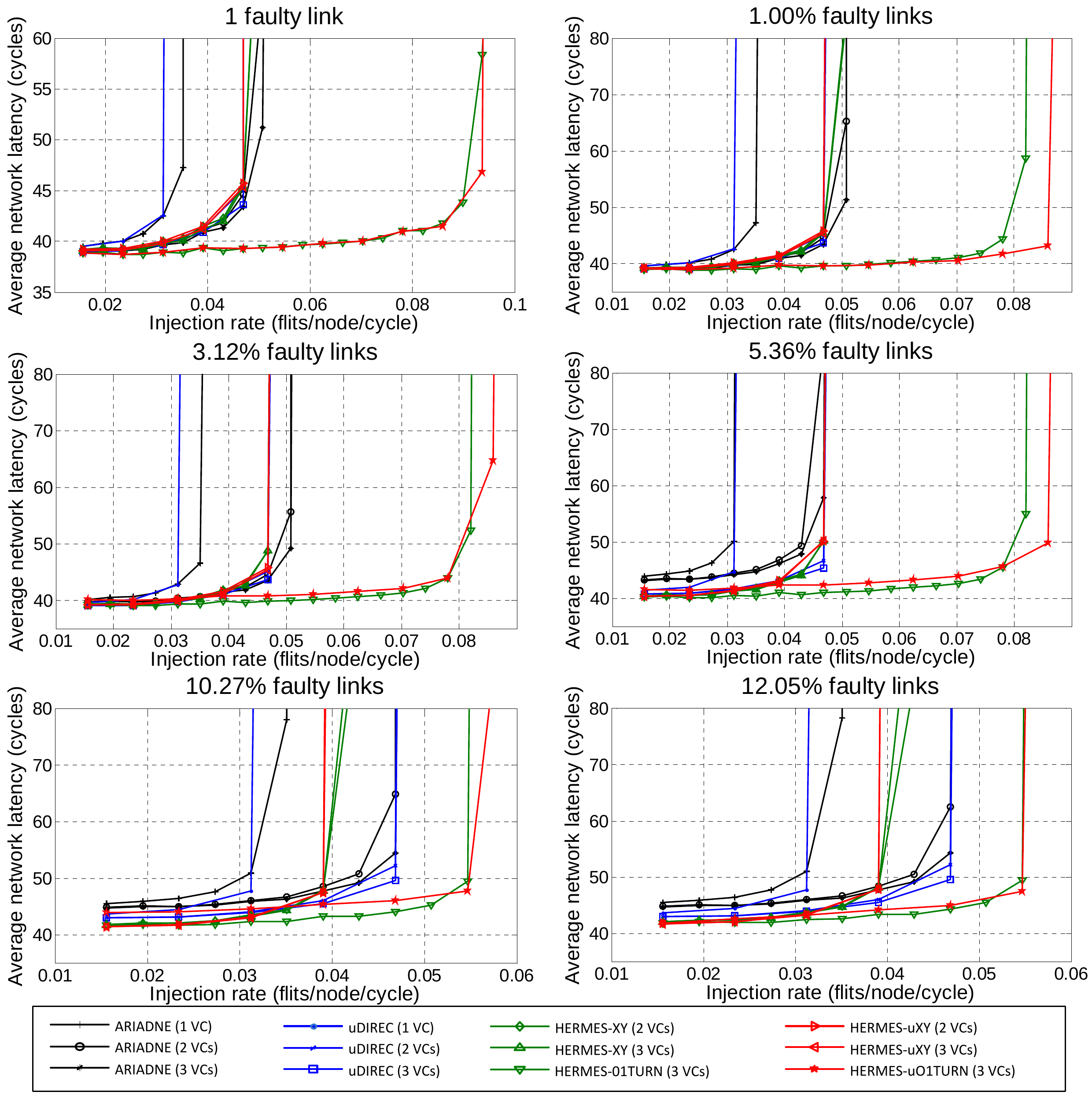}
\vspace{-6mm} \caption{\small{Latency-throughput curves under the
hotspot faulty link placement scenario with synthetic transpose
traffic.}} \label{fig:Hotspot50TransResults}
\end{center}
\vspace{-8mm}
\end{figure}

Fig.~\ref{fig:Hotspot50UniformResults} and
Fig.~\ref{fig:Hotspot50TransResults} show latency-throughput results
under UR traffic and TR traffic respectively, using 1 to 3 VCs per
port for all considered routing schemes. The saturation throughput
of all tested routing algorithms, under each respective faulty link
density scenario, is higher in the case of UR traffic compared to TR
traffic due to said adversarial nature of TR traffic. All uDIREC and
Ariadne variants perform worse here (see
Fig.~\ref{fig:Hotspot50UniformResults}) as compared to their
counterparts under the random faulty links placement using UR
traffic in Section~\ref{subsection:RandomPlaceFaultsRes} and
Fig.~\ref{fig:UniRandPlaceUniRes}, as the concentrated faulty links
in the topology middle cause excessive re-routing equating to
greater hop distances traversed by messages. Interestingly though,
in the case of up to $5.36\%$ faulty links all six Hermes variants
perform better than the equivalent experiments carried out in
Section~\ref{subsection:RandomPlaceFaultsRes} with results shown in
Fig.~\ref{fig:UniRandPlaceUniRes}. This is because the higher
concentration of faulty links in the topology middle leaves fewer
faulty links in the periphery of the hotspot area, which captures
$75\%$ of the topology, in which either XY or O1TURN routing route
efficiently vs. Up*/Down* routing, often utilized in the topology
center, which exhibits reduced path diversity. As such, a higher
chance in using either XY or O1TURN routing arises, reducing the
mean hop distance, hence helping sustain higher network throughput
levels.

Such an advantage is diminished at higher faulty link counts, and in
all experimental setups where TR traffic is used, i.e., when
comparing the results of Fig.~\ref{fig:Hotspot50TransResults} to
those of Fig.~\ref{fig:UniRandPlaceTransRes}, as the combination of
unevenly distributed traffic with the nonuniform concentration of
faulty links creates highly detrimental to performance conditions.
Fig.~\ref{fig:Hotspot50TransResults} shows that all Ariadne and H-XY
variants perform worse for up to $5.36\%$ faulty links versus the
corresponding results of
Section~\ref{subsection:RandomPlaceFaultsRes} and
Fig.~\ref{fig:UniRandPlaceTransRes} where faults are placed randomly.
Also, the performance gap between all the Ariadne and uDIREC variants
with those of all six Hermes variants is quite large, as shown in
Fig.~\ref{fig:Hotspot50TransResults}. H-O1TURN and H-uO1TURN
consistently exhibit the best performance under TR traffic (also for
UR traffic) due to the load-balancing nature of O1TURN routing,
which proves handy in cases where faults are highly
concentrated. Still under TR traffic, at $10.27\%$ and $12.05\%$
faulty link counts the results are slightly smaller, but not too
dissimilar, to those of Fig.~\ref{fig:UniRandPlaceTransRes}, which
shows UR traffic results, for all routing schemes, as due to high
fault numbers all algorithms begin to utilize Up*/Down* routing
heavily (reduced path diversity). With all schemes utilizing 3 VCs
per port and under a single faulty link, H-O1TURN and H-uO1TURN
outperform H-XY and Ariadne by $90.9\%$ and $133.3\%$, and $92.4\%$
and $136.8\%$, respectively, in terms of sustainable throughput,
while H-XY outperforms Ariadne by a ``mere'' $22.2\%$ since H-XY
offers no load-balancing. With a greater number of faults present in
the topology, the performance of H-XY, H-O1TURN, and H-uO1TURN
degrades gracefully. Under the worst-case scenario of  $12.05\%$
faulty links, H-O1TURN's and H-uO1TURN's throughput are $25.0\%$ and
$42.9\%$, and $26.2\%$ and $44.1\%$ higher than those of H-XY and
Ariadne (all with 3 VCs per port), respectively.

\vspace{-2mm}
\subsection{Realistic Full-System Workload Results}\label{subsection:NetraceResults}
\vspace{-1mm}

Realistic workload traces were captured from an $8 \times 8$
mesh-interconnected CMP running all the multithreaded PARSEC v2.1
suite benchmarks~\cite{ParsecBenchRef} executed onto the M5
simulator~\cite{M5Simulator}. The Netrace
infrastructure~\cite{Netrace2010} was used to track dependencies
among packetized messages. Each of the 64 tiles contains an in-order
Alpha core clocked at 2 GHz, each containing separate 32 KB 4-way
set associative L1 I\&D caches with 3-cycle access latency with
coherency maintained via a MESI protocol, and a 16 MB L2 shared
8-way set associative 64-bank fully-shared S-NUCA with 64 B lines
with an 8-cycle access time. Packets consist of 64 bits and 576 bits for
miss request/coherence traffic and cache line transfers,
respectively. All FT mechanisms use 3 VCs per input port, run using
three random faulty link NoC topology placement scenarios: one
faulty link, $5.0\%$ and $10.0\%$ NoC faulty links.

\begin{figure}[t!]
    \begin{center}
        \includegraphics[width=.46\textwidth]{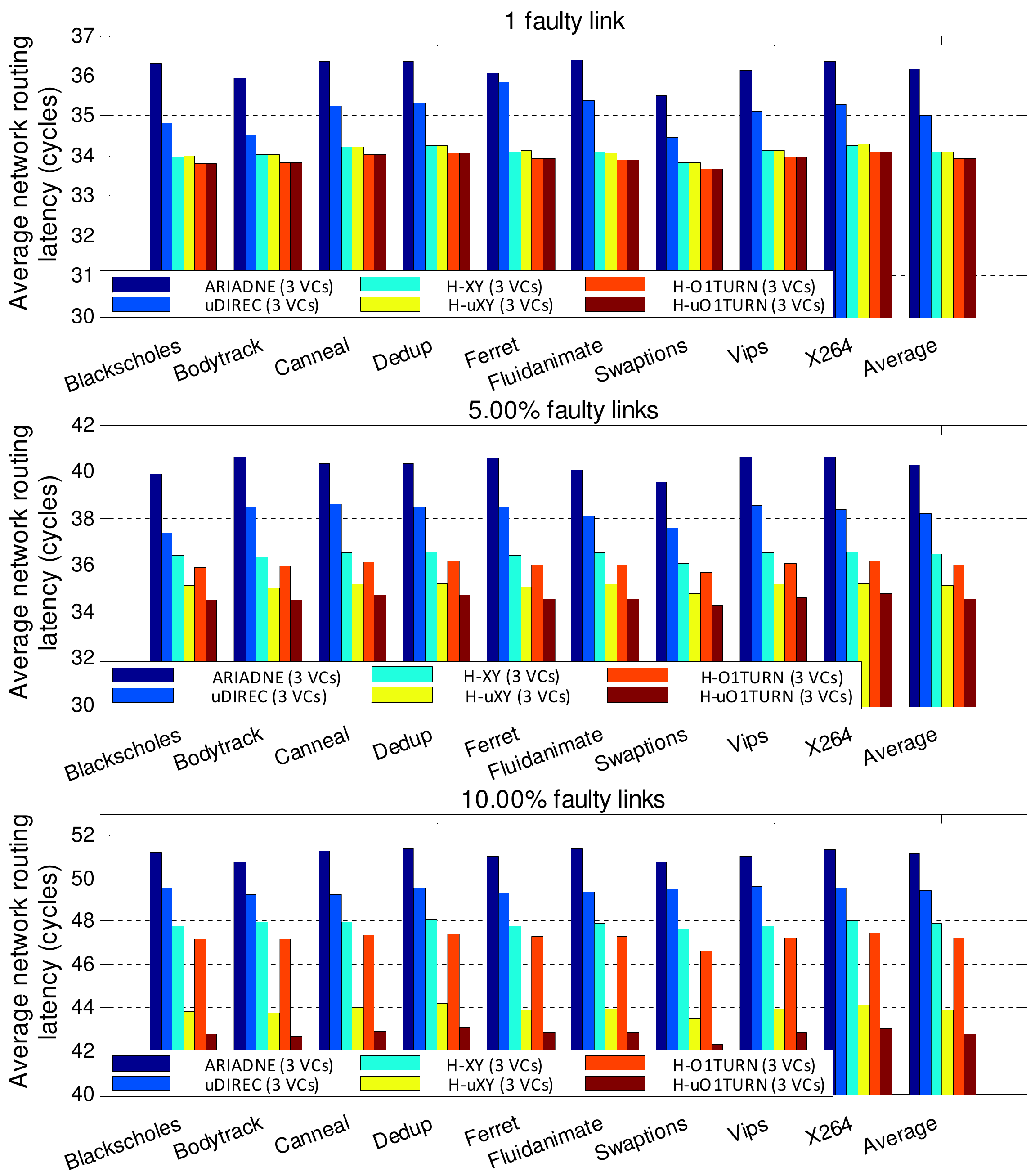}
        \vspace{-3mm}
        \caption{\small{Network routing latency using the Netrace benchmarks.}}
        \label{fig:NetraceRes}
    \end{center}
    \vspace{-7mm}
\end{figure}

Fig.~\ref{fig:NetraceRes} depicts average routing latency results
for all Netrace benchmarks. Ariadne and uDIREC are increasingly
being outperformed by Hermes's four variants with growing faulty
link counts, showcasing  Hermes's superior performance and robustness
attainments. The infusion of the uDIREC scheme in hybrid H-uXY
and H-uO1TURN achieves performance gains against H-XY and
H-O1TURN respectively, due to the enhanced utilization of
unidirectional links and reduced healthy link victimization, with
this gap widening under the severe case of $10\%$ faulty links.
Indicatively, under $5\%$ faulty links, on average H-XY tops Ariadne
and uDIREC by $9.60\%$ and $4.71\%$, H-O1TURN surpasses
Ariadne and uDIREC by $10.66\%$ and $5.83\%$, H-uXY
outperforms H-XY by $3.69\%$  and  H-O1TURN by $2.55\%$, while
H-uO1TURN exceeds H-XY and H-O1TURN by $5.12\%$ and $3.99\%$, all
respectively.

\vspace{-4mm}
\subsection{Zero-Load Latency and Saturation Throughput}\label{subsection:SaturationRes}
\vspace{-1mm}

\begin{figure}[t!]
    \begin{center}
        \includegraphics[width=.49\textwidth]{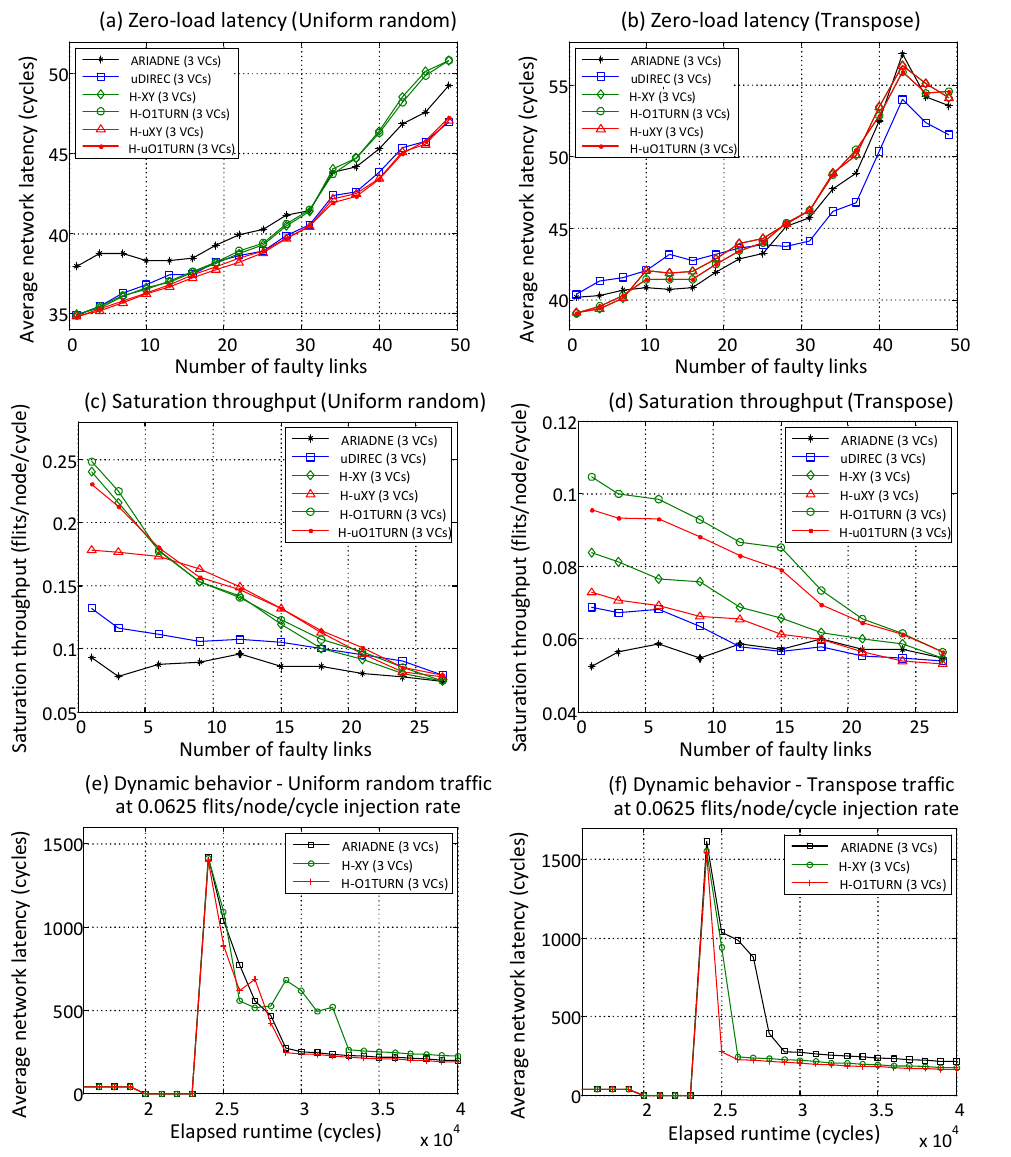}
        \vspace{-7mm}
        \caption{\small{Zero-Load latency under (a) Uniform Random (UR) and (b) Transpose (TR) traffic patterns, saturation throughput with (c) UR and
(d) TR traffic, and network response to dynamic faults: average NoC latency versus time under (e) UR traffic and (f) TR traffic patterns.}}
        \label{fig:SatZeroLDynaRes}
    \end{center}
    \vspace{-8mm}
\end{figure}

An ultra-low injection rate of 0.01 flits/node/cycle is used to
emulate zero-load, where 3 VCs per port are used in all tests.
Fig.~\ref{fig:SatZeroLDynaRes}-(a) shows that under the use of UR
traffic, H-uXY and H-uO1TURN perform identically to uDIREC as the
opportunities for route optimization with O1TURN routing are
non-existent at zero load. All these three schemes, however,
outperform Ariadne, H-XY, and H-O1TURN which victimize a greater
number of healthy links due to their bidirectional use of links vs.
unidirectional under uDIREC; nevertheless, all routing algorithms,
except for Ariadne, perform almost identically for up to $\sim20$
faulty links where a relatively limited number of faulty links is
encountered across an average path traversed. With one faulty link
Ariadne is outperformed by $9.0\%$ by all the other protocols. Under
TR traffic, in Fig.~\ref{fig:SatZeroLDynaRes}-(b), interestingly at
mid to high fault counts, uDIREC outperforms the remaining schemes
due to lesser switching among VCs being carried out when encountering
faults, versus all Hermes variants. With 43 faulty links the
performance of all algorithms begins to improve due to a more
optimal selection of shortest paths picked by the employed Up*/Down*
scheme, despite having fewer healthy links available in the
topology.

Saturation throughput, here, is considered to be the point where the
average NoC latency is $3\times$ its zero-load latency. All tests
carried out utilize 3 VCs per port, with results shown in
Fig.~\ref{fig:SatZeroLDynaRes}-(c) and
Fig.~\ref{fig:SatZeroLDynaRes}-(d) for respective UR and TR traffic
patterns. Under both patterns, Ariadne sustains an almost steady
throughput, exhibiting a small degradation at the highest faulty
links count end, albeit being at the lowest level vs. all other
schemes. This is due to the complete dominance of Up*/Down* routing
which victimizes some links as being bidirectionally faulty,
severely degrading its performance. uDIREC is the second
worst-performing scheme across both UR and TR traffic patterns; it
performs slightly better than Ariadne due to its milder link
victimization under unidirectional Up*/Down* routing. Interestingly,
and consistently, under TR traffic, H-XY outperforms H-uXY, while
H-O1TURN outperforms H-uO1TURN, as under uDIREC, the relatively
heavier reliance on Up*/Down* routing with fewer links being
victimized as faulty, creates elevated congestion vs. alternatively
using XY or O1TURN routing. Overall, under both UR and TR, the four
Hermes variant schemes are performance-superior to Ariadne and
uDIREC. As stated in
Section~\ref{subsection:HermesIntroContributions} the almost
linearly-degrading throughput with an increasing fault links number
(an unavoidable phenomenon) sustained by all Hermes variants is an
indicator of its near-ideal fault-tolerant routing algorithm nature.

\vspace{-4mm}
\subsection{Network Response to Dynamically-Occurring Link Faults}\label{subsection:DynamicRes}

We next evaluate Hermes's dynamic behavior in response to real-time
fault occurrences in NoC links. As such, we set up a scenario where
after an initial stable period of 20,000 cycles, 25 link faults,
randomly distributed in the topology, happen concurrently. At that
point the entire network's packet flow is frozen and no new packets
are network-injected, with flits remaining housed in router
injection buffers. The routing tables are invalidated, and the
reconfiguration process described in
Chapter~\ref{section:HERMESAlgorithm} is then executed. The network
resumes operation after $64^{2}$ ($N^{2}$ routers $\times$ $N^{2}$
cycles/router; $N=8$)= 4,096 cycles (at cycle 24,096 in
Fig.~\ref{fig:SatZeroLDynaRes}-(e,f)), for the assumed 8-ary 2D mesh
NoC, according to timing and syncing requirements of Hermes's
reconfiguration algorithm (refer to
Section~\ref{subsection:TimingSync}).

Fig.~\ref{fig:SatZeroLDynaRes}-(e) and
Fig.~\ref{fig:SatZeroLDynaRes}-(f) show the network reacting to the
aforementioned dynamic link faults occurrences scenario under
respective UR and TR traffic patterns. Each tested routing algorithm
utilizes 3 VCs at each input port, while network latency is measured
at 1,000-cycle intervals. A large spike in network latency starts at
cycle 20,000, due to the network being frozen awaiting for routing
table reconfiguration to complete; this spike is taller under the
highly-uneven TR traffic versus UR traffic. As
Fig.~\ref{fig:SatZeroLDynaRes}-(e,f), shows, under both said traffic
patterns, H-O1TURN takes fewer cycles to stabilize and return to
normal NoC latency levels due to its inherent load-balancing routing
nature and lower mis-routing rate. Under UR traffic, H-XY spends the
longest time to stabilize, due to: (1) the absence of
load-balancing, and (2), 180-degree turning when switching from XY
to Up*/Down* routing which increases packet hop count. Ariadne, on
the other hand, exhibits a more stable behavior since packets follow
pre-determined paths. Under TR traffic, Ariadne consumes the most
cycles to stabilize its behavior, since it does not feature
load-balancing capability, with only one unique path provided
between each source-destination router pair, as defined by Up*/Down*
routing. As such, these result to higher contention and consequent
routing delays across available routing paths.

\vspace{-3mm}
\subsection{Hardware Synthesis Results}\label{subsection:Synthesis}
\vspace{-1mm}

We implemented and synthesized \textit{all the Hermes hardware
blocks} illustrated in Fig.~\ref{fig:HermesRouterArch} and described
in Section~\ref{section:HERMESArch} using the Synopsys Design
Compiler, targeting a commercial 45 nm CMOS technology library at 1
V. We consider three-stage speculative pipelined virtual-channel
routers~\cite{PehDelayModel} with credit-based wormhole flow-control
with a 64-entry SRAM-based routing table, 128-bit links, 6-flit
buffers, 2 GHz clock rate, at $50\%$ switching activity, all for an
$8 \times 8$ mesh topology. The 2-bit overlay network described in
Section~\ref{section:HERMESAlgorithm} and
Section~\ref{section:HERMESArch} possesses Triple Modular Redundancy
(TMR). Ariadne~\cite{Aisopos2011} was also implemented, however the
uDIREC reconfiguration scheme~\cite{uDIRECpaper} is not compared
against Hermes as it is heavily dependent on software strategies to
form routing paths; in addition, the authors do not specify the
implementation of the required end-to-end ECC blocks, and only
report area overheads. It is, however expected, that once the
reconfiguration hardware, the routing tables and ECC blocks are
taken into account, that uDIREC will surpass Hermes's overheads.

Table~\ref{table:SynthRes} briefs power-area overheads for the base
speculative NoC router, as well as Ariadne, H-XY and H-O1TURN with 1
up to 3 VCs per input port. H-XY (2 VCs/port) presents $5.31\%$ area
and $9.81\%$ power overheads respectively as compared to the base 2
VC per-port NoC router, while the corresponding comparisons against
Ariadne with 2 VCs are $0.94\%$ and $4.95\%$. H-O1TURN (3 VCs/port)
presents $4.47\%$ area and $5.91\%$ power overheads respectively as
compared to the base 3 VC per-port NoC router, while the analogous
comparisons against Ariadne with 3 VCs are $1.51\%$ and $0.47\%$.

\vspace{-2mm}
\section{Background and Related Work}\label{section:RelatedWork}
\vspace{-1mm}

Fault-Tolerant (FT) approaches applicable to NoCs have been inspired
from macro-level Interconnection Networks (INs), where Radetzki's
survey~\cite{RadetzkiFTMethods}, and Dally's and Duato's
textbooks~\cite{DallyBook,DuatoBook}, provide a broad
coverage for each respective domain. The scope of any FT approach
is universal: to sustain seamless communication among all
interconnected entities in the presence of \textit{faulty links},
\textit{faulty nodes}, or \textit{faulty regions}. Duato's landmark
work~\cite{DuatoFTTheory} is conducive in developing a theory for FT
routing in INs, and consequently NoCs. Essentially, as long as FT
routing provides full connectivity devoid of cyclic channel
dependencies in a sub-connected (i.e., faulty) topology, then the FT
function crucially guarantees deadlock- and livelock-freedom during
packet delivery.

Most FT approaches are categorized as: (1) FT routing algorithms
(FTRAs) that bypass link/router failures, (2) logic/architectural
redundancy within routers to improve their resilience, and, (3)
hybrid approaches that combine (1) and (2). Category (1) is further
sub-classified as: (a) FTRAs with bounded FT support, (b) FTRAs that
sustain unbounded faulty link counts but with spatial pattern
limitations, and, (c) FTRAs that bolster unbounded faulty link
counts and no spatial placement restrictions. FT schemes in the
latter category, e.g.,  Hermes, are the most flexible, albeit
challenging to devise.

The restrictive FT approaches outlined in categories (1-a) and (1-b)
above dominated the initial research attempts in INs. Under (1-a),
Dally's early reliable router is a 1-FT
architecture~\cite{RelRouterDally}, while the  interested reader is
urged to refer to related early works found
in~\cite{DallyBook,DuatoBook} for holistic coverage. Under category
(1-b), in an effort to define permissible spatial fault patterns to
avoid deadlocks and simplify FT routing, researchers utilized
block-structured shapes, such as convex and/or concave regions of
faults, at the expense of victimizing healthy links and
routers~\cite{FTOddEven}. Under category
(1-c),~\cite{ResilientRoute} proposes distributed routing algorithms
that re-configure a NoC to avoid faulty components. Next, the Vicis
router~\cite{VicisGates} employs special BIST testers to detect
faults, and then leverages extensive re-configurability and an
appropriately-designed FTRA to perform port-swapping and crossbar
bypassing.

As Hermes  reconfigures a faulty topology using Up*/Down* (UD)
rules, we next outline the historical timeline of UD routing so as
to delineate its evolution. The fundamental work by
Perlman~\cite{PerlmanSpanning} presented a self-configuring
 algorithm that maintains an acyclic spanning
subset of a general mesh topology that converges in time
proportional to the diameter of its extended LAN. Next,
Autonet~\cite{AutoNetPaper}, which comprises a self-configuring LAN,
utilizes a distributed algorithm running on switch processors to
compute new routes when extra links are incorporated or when
existing links fail. Deadlock-free routing and the flooding pattern
for broadcast packets in Autonet are both based on computing a
spanning tree of operational links inspired from principles outlined
in Perlman's algorithm, and extended with the use of the then
proposed UD routing which performs BFS from a root node that
establishes deadlock freedom and full network connectivity. Next,
work by Sancho~\emph{et
al.}~\cite{SanchoNewMethodNOW,Sancho2004TPDS} proposed the use of UD
routing in NoWs with irregular topologies where deadlock-free
routing tables are computed based on Depth-First Search (DFS;
instead of BFS) spanning tree generation to eliminate cyclic
dependencies where throughput is improved with the use of minimal
routes and traffic balancing, albeit with no support for fault
tolerance. The same authors also proposed the use of a Flexible
Routing scheme~\cite{SanchoFlex} applicable to regular networks,
i.e., meshes. Following, Silla~\emph{et al.}~\cite{SillaPaper}
extend the original UD topology-agonistic routing in irregular NoWs
and proposed two general methodologies toward the design of
high-performance adaptive routing algorithms, albeit with no fault
tolerance support.

\begin{table}[t!]
    \caption{\small{Overheads of various synthesized NoC routers.}}
    \begin{center}
        \vspace{-2mm}
        \includegraphics[width=.43\textwidth]{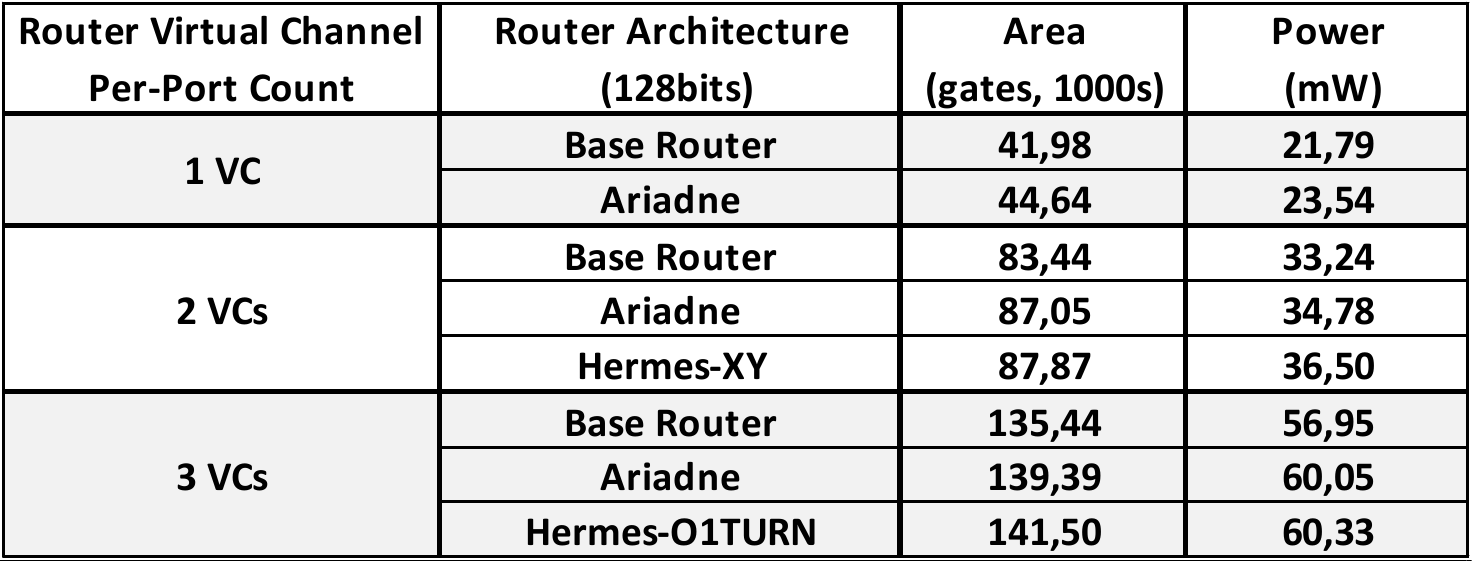}
        \label{table:SynthRes}
    \end{center}
    \vspace{-8mm}
\end{table}

Topology-agnostic algorithms that discover deadlock-free routes via
reconfiguration have also become handy for faulty NoCs. Ariadne's
reconfiguration algorithm~\cite{Aisopos2011} which uses UD routing
to compute routes when faults in links occur was first devised.
Next, uDIREC~\cite{uDIRECpaper} reconfigures routing paths in a
faulty NoC by also using UD routing, where an iterative
software-based spanning tree kernel utilizes NoC path diversity to
optimize routes with restricted unidirectional link victimization
vs. Ariadne. Last, Fault- and Application-aware Turn model Extension
(FATE) routing~\cite{DLeeBertaccoICCD15} also utilizes a
topology-agnostic iterative-based software approach to generate a
deadlock-free application-aware routing function.

Under category (2), prominently, Bulletproof~\cite{BulletProof}
analyzes the reliability vs. area tradeoffs of various NoC router
designs and proposes run-time repair and recovery methodologies at
the system-level. Next,~\cite{Vitkovskiy2012} makes use of partially
faulty links, while~\cite{LinkReversal} converts a uni-directional
link to work in a full-duplex mode in case the node-pairing link
fails. Last, under category (3), stochastic communication
techniques~\cite{Marculescu03a} use probabilistic packet
broadcasting schemes to handle faults.

\vspace{-3mm}
\section{Conclusions}\label{section:conclusions}
\vspace{-1mm}

This paper presented Hermes, a near-ideal, high-throughput,
distributed, and deadlock-free FT routing algorithm with high
robustness and graceful performance degradation with increasing
faulty link counts. Hermes is a hybrid routing scheme: it balances
traffic to sustain high performance onto fault-free paths, while it
provides pre-configured escape path selection in the vicinity of
faults. Hermes improves throughput by up to $3\times$, and is able
to identify network segmentations.

\vspace{-14mm}
\begin{IEEEbiography}[{\includegraphics[width=1in,clip,keepaspectratio]{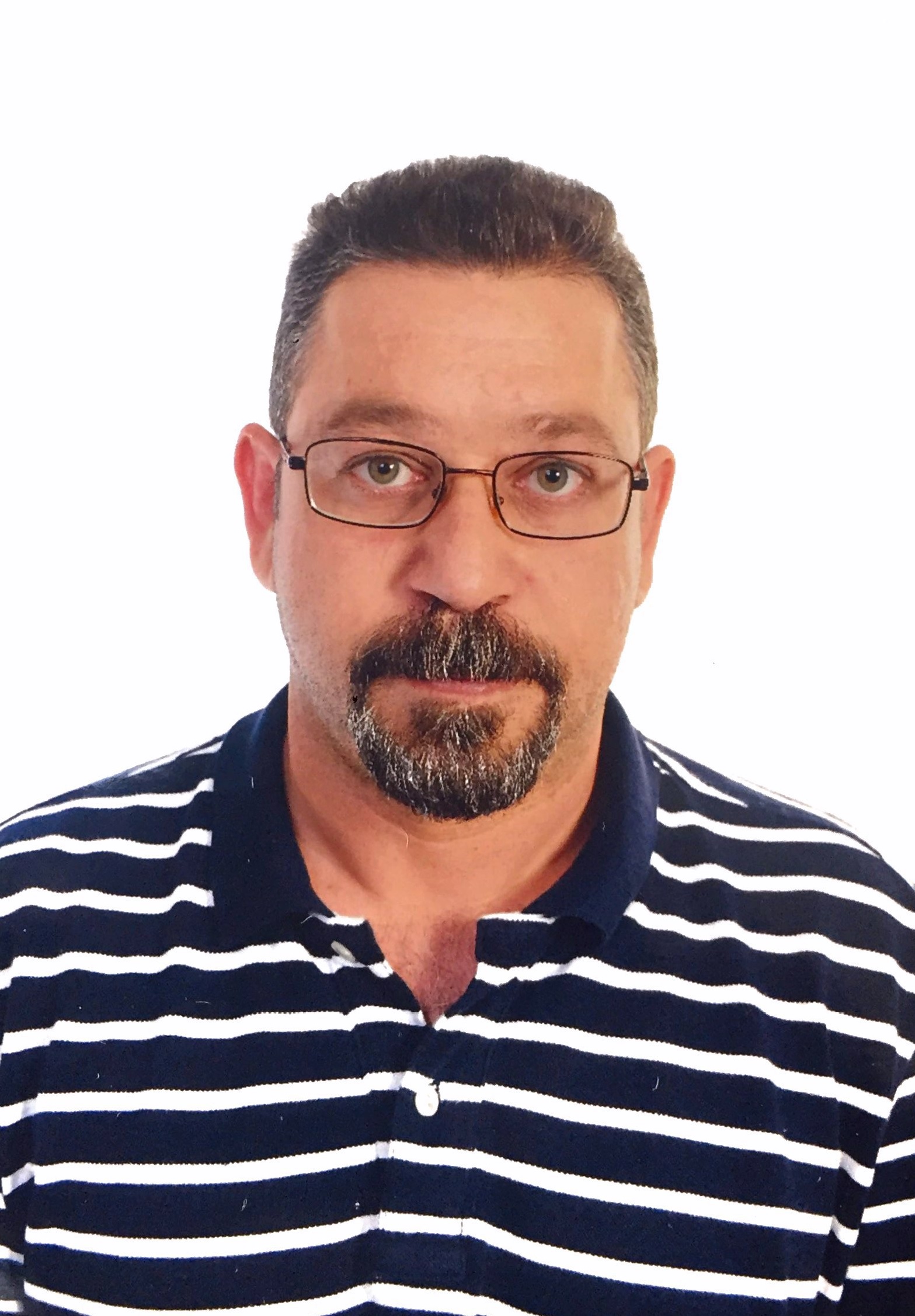}}]{Costas Iordanou} received the B.Sc.
(2013) and M.Sc. (2014) degrees in Computer Engineering and Informatics from the Cyprus University of Technology. He also holds a M.Sc. degree in
Telematic Engineering from  Universidad Carlos III de Madrid (2015). Currently he is working on completing his Ph.D. degree (Marie Curie, ITN-METRICS
program) at the Universidad Carlos III de Madrid in collaboration with Telefonica I + D. His research interests focus on targeted web advertising, web
tracking, online price discrimination, and fault-tolerant on-chip networks.
\end{IEEEbiography}
\vspace{-14mm}

\begin{IEEEbiography}[{\includegraphics[width=1in,clip,keepaspectratio]{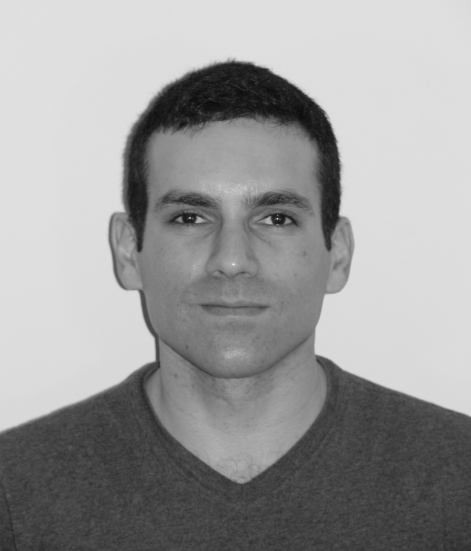}}]{Vassos Soteriou} received the
B.S. and Ph.D. degrees in electrical engineering from Rice University, Houston, TX, in 2001, and Princeton University, Princeton, NJ, in 2006,
respectively. He is currently an Associate Professor at the Department of Electrical Engineering, Computer Engineering and Informatics at the Cyprus
University of Technology. He is a recipient of a Best Paper Award at the 2004 IEEE International Conference on Computer Design. His research interests lie
in multicore computer architectures, and on-chip networks.
\end{IEEEbiography}
\vspace{-14mm}

\begin{IEEEbiography}[{\includegraphics[width=1in,clip,keepaspectratio]{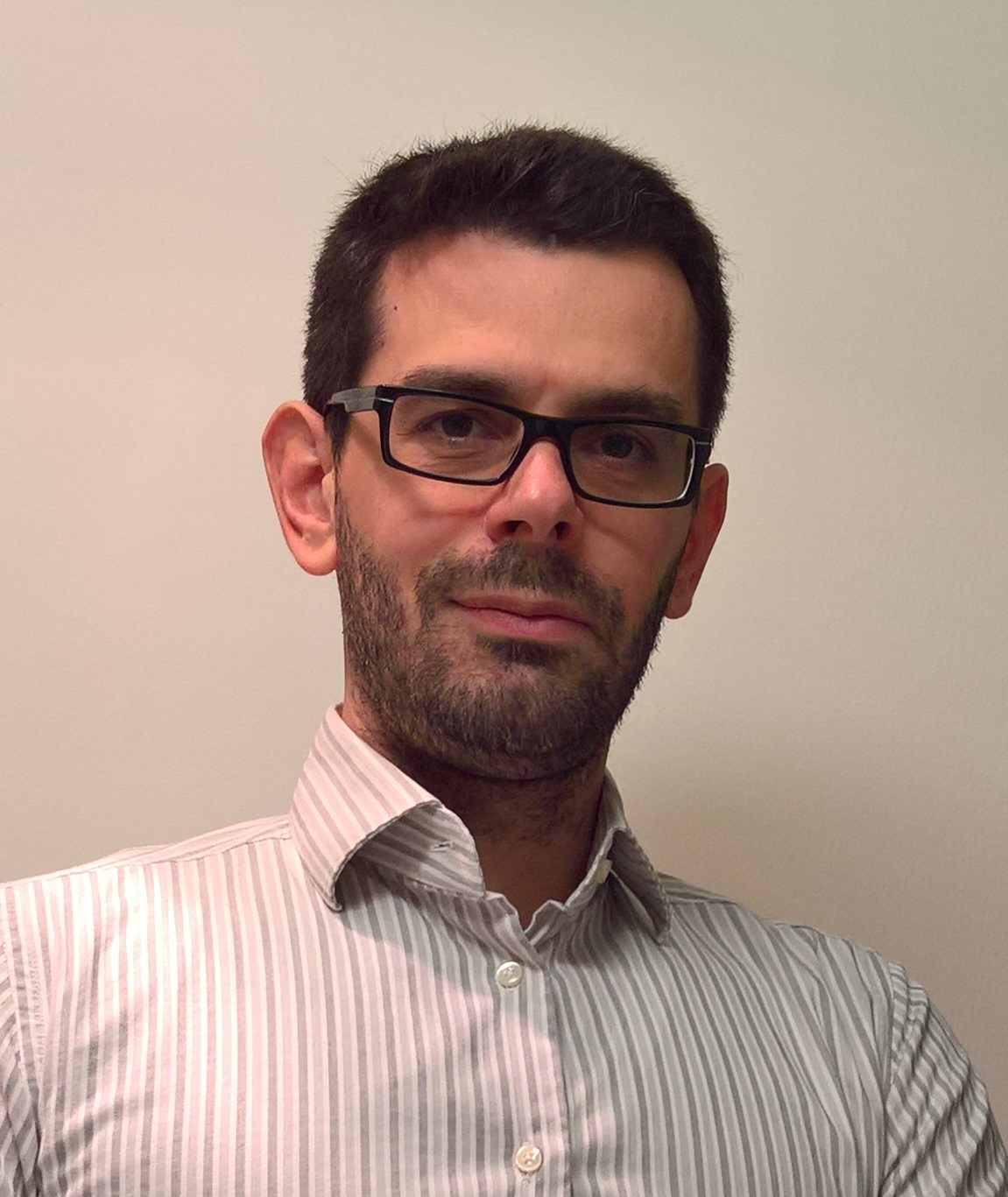}}]{Konstantinos Aisopos} received his PhD degree from Princeton University in 2012. His research has been on on-chip networks and reliability. During his Ph.D. studies, he spent
three years at MIT, developing a circuit-level accurate fault-modelling tool for on-chip routers. He then spent five years at Microsoft Seattle, working
on Cortana and developing an artificially intelligent digital assistant for Windows. During 2017-2019, he co-founded a
startup on Machine Learning in Davis, CA. Since 2019 he has been working at Google in Mountain View CA as a senior software engineer.
\end{IEEEbiography}

\end{document}